\documentclass{aa}
\usepackage{txfonts}
\usepackage{epsfig}
\usepackage{lscape}
\bibpunct{(}{)}{;}{a}{}{,}    
\usepackage{graphicx,graphics}
\usepackage{color}
\usepackage{ulem}
\usepackage{float}
\usepackage{longtable}
\usepackage{amsmath}
\usepackage{amssymb}
\usepackage{adjustbox}
\usepackage[breaklinks=true, colorlinks=true, urlcolor=blue, citecolor=blue, linkcolor=blue]{hyperref}

\begin{document}

\title{Electron temperature relations in star-forming regions}
\subtitle{I. The ionization-dependent $t_2$--$t_3$ relation}

\author{
        I.~A.~Zinchenko\inst{\ref{ARI}} \and
        B.~P\'{e}rez-D\'{i}az\inst{\ref{INAF}} \and
        E.~P\'{e}rez-Montero\inst{\ref{IAA}} \and
        J.~M.~V\'{i}lchez\inst{\ref{IAA}}
}
       
\institute{
Astronomisches Rechen-Institut, Zentrum f\"{u}r Astronomie der Universit\"{a}t Heidelberg, M\"{o}nchhofstra{\ss}e 12-14, D-69120 Heidelberg, Germany\label{ARI}
\and
INAF - Osservatorio Astronomico di Roma, via Frascati 33, I-00078 Monte Porzio Catone, Italy \label{INAF}
\and
Instituto de Astrof\'{i}sica de Andaluc\'{i}a (CSIC), Apartado 3004, 18080 Granada, Spain \label{IAA} 
}

\abstract
{The relation between the electron temperatures in the high- and low-ionization zones of \ion{H}{ii} regions (the $t_2$--$t_3$ relation) is a key assumption of the direct ($T_{\rm e}$) method of gas-phase abundance determination. Classical parameterizations assume a one-dimensional correspondence between $t_2$ and $t_3$, but they were calibrated over a limited range of nebular conditions and may not be applicable to the extreme ionization and metallicity regimes probed by modern observations, including high-redshift star-forming galaxies.
}
{We investigate the nonlinearity and scatter in nebular electron-temperature relations and provide a parameterization applicable across the conditions typical for local and high-redshift star-forming regions.
}
{We computed a large homogeneous grid of photoionization models with \textsc{Cloudy} C23.01, covering $12+\log({\rm O/H})=6.8$--9.2, $\log U=-4.0$ to $-1.2$, $\log({\rm N/O})=-2.0$ to 0.0, electron densities of 100, 500, and 1000 cm$^{-3}$, ionizing-source ages of 1--6 Myr, and both POPSTAR and BPASS stellar spectral energy distributions. We compared ion-weighted temperatures with temperatures inferred from auroral-to-nebular line ratios and parameterized the $t_2$--$t_3$ relation using one- and two-dimensional polynomial fits.
}
{The $t_2$--$t_3$ relation predicted by our grid is nonlinear, with the dominant dependencies on the ionization parameter and the hardness of the ionizing radiation field while the electron density plays only a minor role. 
A two-dimensional parameterization by $\log({\rm O32})$ significantly reduces the residual scatter from $\sim3$--4\% for the one-dimensional relation to $\sim1$--2\%. The same fit with S2S3 in place of O32 matches it below $T_{\rm e}({\rm O}^{++})=14\,000$~K and is less accurate above. Electron density contributes little additional predictive power. Dust introduces a systematic shift in the calibration that becomes significant only at high temperatures. Temperatures derived from auroral-to-nebular line ratios agree with ion-weighted temperatures to within a few percent over most of the grid, but they diverge by up to $\sim20\%$ in the most extreme models, where low temperatures coincide with high excitation. Among the considered diagnostics, $T_{\rm e}([\ion{N}{ii}])$ and $T_{\rm e}([\ion{S}{iii}])$ provide reliable tracers of $T_{\rm e}({\rm O}^{+})$ and $T_{\rm e}({\rm O}^{++})$, respectively.
}
{Classical one-dimensional $t_2$--$t_3$ relations do not completely describe the full range of nebular conditions explored by modern photoionization models. Their use can introduce systematic chemical abundance biases, particularly in the high-ionization, high-temperature regime typical for many observed metal-poor and high-redshift galaxies. The ionization-dependent relations presented here provide a physically motivated framework for reducing these biases and interpreting direct-method abundances across a broad range of nebular conditions. They are derived for individual idealized nebulae, and their application to spatially integrated galaxy spectra still requires observational validation.
}


\titlerunning{Electron temperature relations in star-forming regions}
\authorrunning{Zinchenko et~al.}
\maketitle


\section{Introduction}

Chemical abundances in \ion{H}{II} regions are commonly determined using the direct, or $T_{\rm e}$, method. This approach relies on measuring electron temperature from the ratio of strong nebular lines such as [\ion{O}{III}]$\lambda$5007, [\ion{N}{II}]$\lambda$6584, [\ion{S}{III}]$\lambda$9069, and [\ion{O}{II}]$\lambda\lambda$3727,3729 and corresponding faint auroral lines such as [\ion{O}{III}]$\lambda$4363, [\ion{N}{II}]$\lambda$5755, [\ion{S}{III}]$\lambda$6312, and [\ion{O}{II}]$\lambda\lambda$7320,7330. However, H\,\textsc{ii} regions are not isothermal and different ions dominate at different depths within the nebula because photons of different energies penetrate different distances before being absorbed. Consequently, O$^{+}$ is found predominantly in the outer, lower ionization zone, whereas O$^{++}$ occupies the inner, higher ionization zone. Once the electron temperature and density in a given ionization zone are known, ionic abundances can be derived directly from nebular lines using statistical equilibrium calculations \citep[e.g.,][]{Osterbrock2006,PerezMontero2017}. 
In practice, however, only a subset of auroral lines is usually detected. This creates a fundamental difficulty since the temperatures characterizing the low- and high-ionization zones of the nebula are generally not measured simultaneously. Instead, one commonly derives the temperature of the ionization zone from the other through an assumed relation between the temperature in low ionization zone $t_2$ and the temperature in high ionization zone $t_3$. Here, $t_2$ and $t_3$ are the temperatures of the O$^{+}$ and O$^{++}$ zones.

Classical $t_2$--$t_3$ relations were derived from early photoionization models using the stellar atmosphere models and atomic data available at the time and calibrated over relatively limited ranges of nebular conditions \citep{Stasinska1980,Campbell1986,Pagel1992,Garnett1992}. These relations typically adopted a one-dimensional correlation between $t_2$ and $t_3$. Such parameterizations became standard tools in direct method abundance studies and remain widely used today.

Subsequent theoretical and observational studies, however, have shown that the relation may be more complex. Photoionization models suggested departures from linearity and possible dependencies on metallicity, excitation, or density \citep{Oey2000,PerezMontero2003,Izotov2006}. The dependence of the dispersion on electron density was discussed explicitly with models by \citet{Hagele2006} and is summarized in \citet{PerezMontero2017}, while \citet{PerezMontero2014} provided a $t_2$--$t_3$ relation based on the {\sc HII-CHI-mistry} photoionization grid. Observational studies further showed that \ion{H}{II} regions with similar $t_3$ values can exhibit systematically different $t_2$ temperatures depending on the excitation state and ionization conditions \citep{Pilyugin2007,Pilyugin2009,ArellanoCordova2020}. More recent analyses have emphasized the importance of internal temperature structure, density inhomogeneities, and shocks in shaping observable temperature diagnostics \citep{MendezDelgado2023,RickardsVaught2024,RickardsVaught2025}.

At the same time, James Webb Space Telescope (\textit{JWST}) NIRSpec spectroscopy is extending auroral-line studies to galaxies at cosmic noon and into the epoch of reionization. These systems frequently occupy regions of parameter space characterized by a high ionization parameter, hard radiation field, and extreme excitation conditions that are only sparsely represented in local calibrating samples \citep[e.g.,][]{Cataldi2025}. An important question then is how far the standard one-dimensional $t_2$--$t_3$ relations remain applicable under such conditions.

A related issue is that the temperatures derived observationally from auroral-to-nebular line ratios are not identical to the ion-weighted temperatures predicted by photoionization models. Because auroral transitions arise from levels with substantially higher excitation energies than the corresponding nebular transitions, their emissivities are much more sensitive to electron temperature. Consequently, auroral-to-nebular line ratios preferentially weight the hottest regions of the nebula and may systematically differ from the underlying physical thermal structure \citep{Peimbert1967,Peimbert2017}. Distinguishing between physical and observable temperatures is therefore essential when interpreting temperature relations and their impact on chemical abundance determinations.

In this work we study the $t_2$--$t_3$ relation using a large homogeneous grid of \textsc{Cloudy} C23.01 photoionization models spanning a broad range of metallicity, ionization parameter, N/O ratio, density, and stellar-population age. Our goal is to investigate the physical origin of the nonlinearity and scatter in the temperature relation, identify the dominant secondary parameters, and derive practical parameterizations applicable across the nebular conditions sampled by nearby and high-redshift star-forming galaxies.

The paper is organized as follows. Section~\ref{sect:grid} describes the model grid and numerical setup. Section~\ref{sect:temperature} compares ion-weighted and line-derived temperatures, investigates the structure of the $t_2$--$t_3$ relation, and derives one- and two-dimensional parameterizations. Section~\ref{section:Discussion} discusses the implications and limitations of the results in the context of previous theoretical and observational work. Section~\ref{section:Summary} summarizes the main conclusions.

\begin{figure*}
\resizebox{1.00\hsize}{!}{\includegraphics[angle=000]{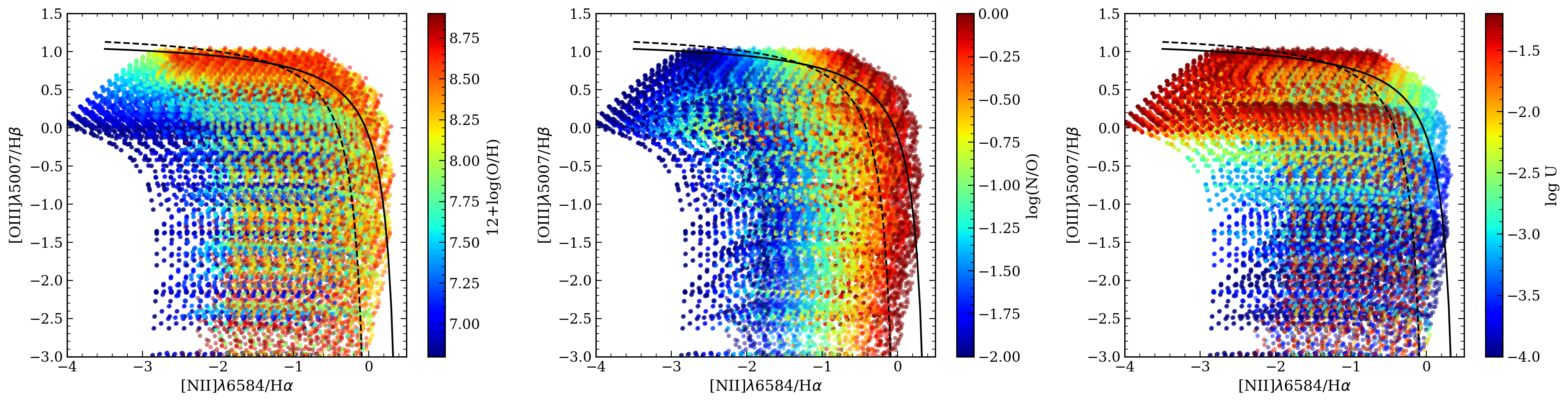}}
\caption{%
BPT diagram for the full model grid (ionizing-source ages 1--6~Myr, densities 100--1000~cm$^{-3}$). Three panels present the same data set color-coded with respect to oxygen abundance, N/O ratio, and $\log \rm{U}$. The solid line is the theoretical upper limit for \ion{H}{II} regions presented by \citet{Kewley2001}. The dashed line is the demarcation line proposed by \citet{Kauffmann2003} as a lower limit for the area of AGNs.
}
\label{figure:BPT}
\end{figure*}

\section{The model grid}
\label{sect:grid}

\subsection{Grid definition}
\label{sect:grid-def}

A large grid of photoionization models was calculated to obtain emission line fluxes of different chemical elements and electron temperatures in corresponding ionization zones. The models were computed with the \textsc{Cloudy} C23.01 code \citep{CloudyC23}, assuming a plane-parallel geometry and a constant electron density of 100, 500, and 1000~cm$^{-3}$. The spectral energy distribution (SED) of the ionizing source was taken from the POPSTAR models \citep{POPSTAR2009} for a \citet{Chabrier2003} stellar initial mass function, with an instantaneous starburst of age 1, 2, 3, 4, 5, and 6~Myr and the same metallicity as the gas. In addition, similar set of grids, also covering ages of 1--6~Myr, was computed with BPASS v2.3 binary and single-star SEDs \citep{BPASS23_p2,BPASS23_p1}, with the initial mass function (IMF) close to the POPSTAR \citet{Chabrier2003} IMF except in its upper mass limit of 100~$M_{\odot}$. Photoionization grids computed with BPASS SEDs are used to assess the dependence of the temperature relation on the shape of ionizing SED.

We varied the oxygen abundance $12+\log$(O/H) from 6.8 to 9.2 and the ionization parameter $\log U$ from $-4.0$ to $-1.2$, while $\log$(N/O) was varied from $-2.0$ to $0.0$. All other chemical abundances were scaled in solar proportions to O, which was assumed to be $12+\log(\mathrm{O/H})_{\odot}=8.69$ \citep{Asplund2009}. Nitrogen is the one element decoupled from this scaling, so that the nitrogen lines can be explored without imposing a fixed N/O--O/H relation, following the strategy adopted for the {\sc HII-CHI-mistry} grids \citep{PerezMontero2014}. In total, about 2200 combinations of the metallicity, $\log U$, and $\log$(N/O) were computed for each age and density of the star-forming region. The remaining modeling choices, common to all the grids analyzed here, are as follows. The filling factor is 0.1. Calculations are iterated four times and stopped at the electron-fraction criterion $\mathrm{efrac}=0.01$ or at the Cloudy default minimum temperature of 4000~K, whichever is reached first. No cosmic-ray heating or extragalactic-background ionizing field is added on top of the stellar SED.  
Adding cosmic-ray heating corresponding to the default Galactic background cosmic-ray ionization rate to a representative subset of models changed the derived electron temperatures by less than 0.02\%, confirming that its effect is negligible for the present calculations.

Compared with the photoionization grids underlying the classical $t_2$--$t_3$ relations \citep[e.g.,][]{Stasinska1980,Pagel1992,Garnett1992,Izotov2006}, the grid presented here differs from earlier ones in two main respects. First, \textsc{Cloudy} C23.01 includes the latest updates of the collisional and recombination atomic data relevant for the nebular and auroral lines used in $T_{\rm e}$ diagnostics. Second, the POPSTAR ionizing SEDs include the contribution of Wolf-Rayet (WR) stars at intermediate ages ($\sim 3$--$5$~Myr), allowing us to assess explicitly the impact of the hardness of the ionizing radiation field on the thermal structure of the nebula.

In Fig.~\ref{figure:BPT} we present the distribution of the models from our grid on the $\log$([\ion{O}{III}]$\lambda$5007/H$\beta$) versus $\log$([\ion{N}{II}]$\lambda$6584/H$\alpha$) BPT diagram~\citep{BPT}. Most models reproduce the locus of \ion{H}{II} regions below the empirical \citet{Kauffmann2003} and theoretical \citet{Kewley2001} demarcation lines. Nevertheless, models with high nitrogen enrichment, log(N/O) $\gtrsim -0.5$, tend to move toward the composite region between the two demarcation lines due to their enhanced [\ion{N}{II}]/H$\alpha$ ratios. The combination of high N/O and high ionization parameter further drives some models above the \citet{Kewley2001} line into the nominal active galactic nucleus (AGN) region, a displacement already described for nitrogen-enriched star-forming models by \citet{PerezMontero2009b}. We retain these models because they represent N-enhanced photoionized nebulae rather than AGNs.

\subsection{Numerical convergence and parameter sensitivity}
\label{sect:grid-conv}

Because this work requires accurate emission-line intensities, including faint auroral lines, across a wide range of nebular conditions, we verified that the grid is numerically converged and robust against the main modeling choices. Convergence was assessed by comparing the relative differences of emission-line intensities between grids computed with different numerical parameters: the electron-fraction stopping criterion {\it efrac}, the number of {\it iterations}, and the temperature-convergence tolerance. The numerical parameters change the line intensities, and the derived electron temperatures, by well below 1\%.

Structural assumptions produce a larger, though still generally modest, impact on the predicted emission-line intensities. Doubling the filling factor has a negligible effect, whereas switching between plane-parallel and spherical geometry can alter individual line intensities by up to $\sim 10$\%. However, auroral and nebular lines from the same ion respond in a similar manner, making nebular-to-auroral ratios substantially less sensitive to the adopted geometry. The dust content is analyzed separately in Sect.~\ref{sect:dust}.

\begin{figure*}
\resizebox{1.00\hsize}{!}{\includegraphics[angle=000]{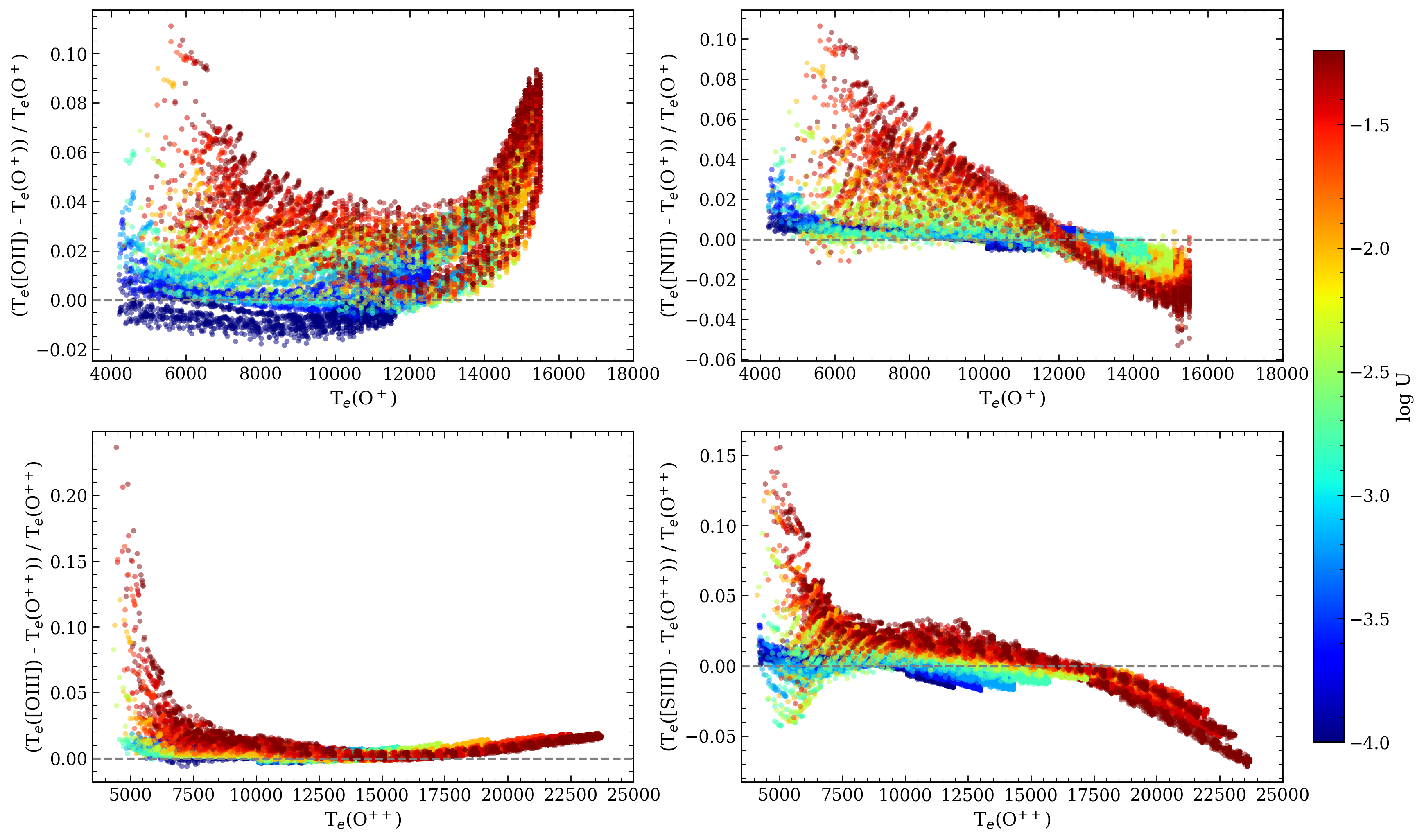}}
\caption{Relative differences between line-derived and ion-weighted electron temperatures versus the ion-weighted temperature for the 1--3~Myr models (before the Wolf-Rayet phase). Panels: T$_e$([\ion{O}{II}]) vs T$_e$(O$^+$), T$_e$([\ion{N}{II}]) vs T$_e$(O$^+$), T$_e$([\ion{O}{III}]) vs T$_e$(O$^{++}$), and T$_e$([\ion{S}{III}]) vs T$_e$(O$^{++}$). Points are color-coded by $\log U$. The dashed line marks zero difference.}
\label{figure:Te-comparison-young}
\end{figure*}

\begin{figure*}
\resizebox{1.00\hsize}{!}{\includegraphics[angle=000]{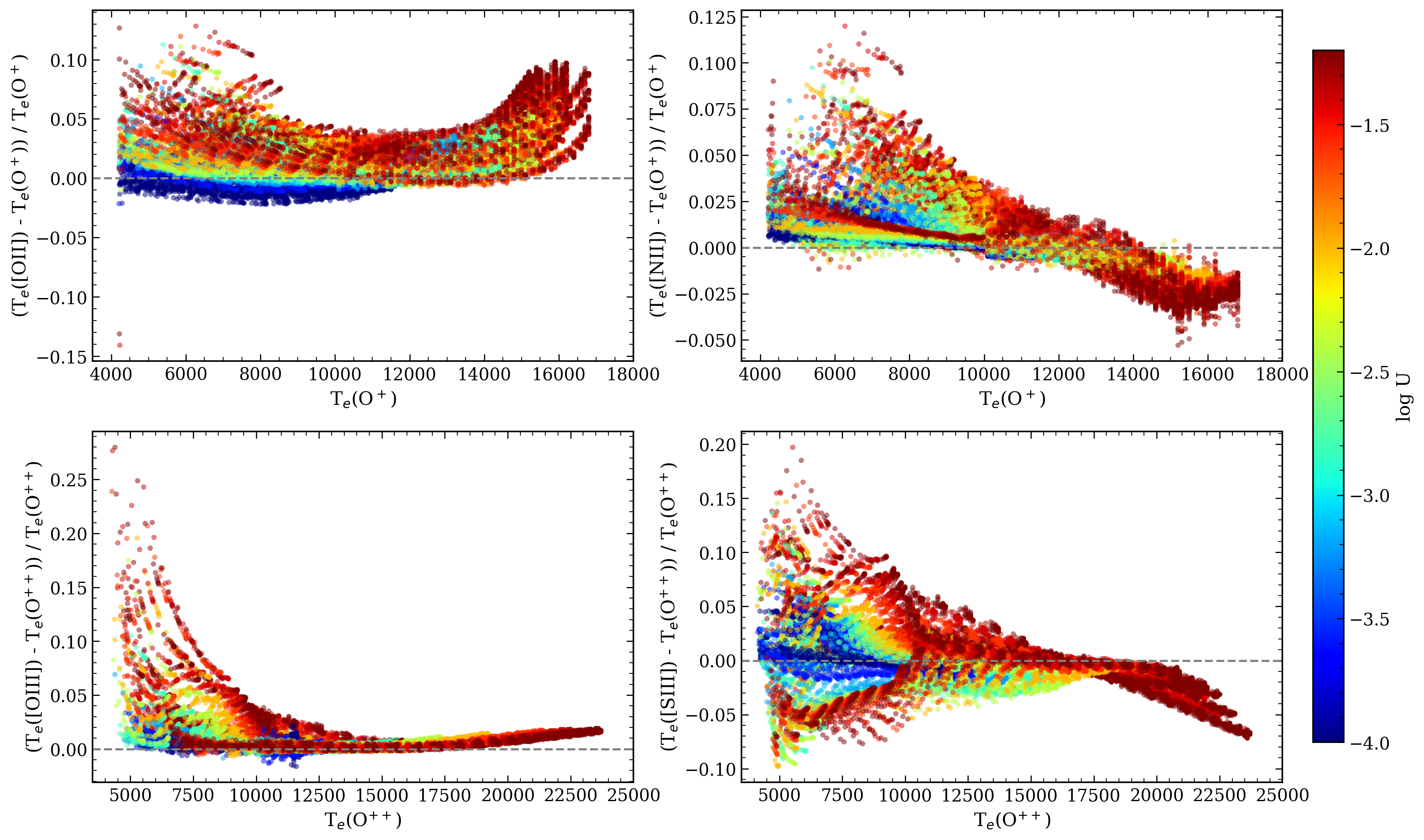}}
\caption{As Fig.~\ref{figure:Te-comparison-young} but for the full model grid spanning ionizing-source ages of 1--6~Myr.}
\label{figure:Te-comparison-old}
\end{figure*}

\section{Electron temperatures}
\label{sect:temperature}

When computing our model grid, the \textsc{Cloudy} outputs were normalized to $I(\mathrm{H}\beta)=100$, allowing emission lines as faint as 0.0001\% of H$\beta$ to be retained in the exported output.
This allows electron temperatures to be derived from the predicted auroral-to-nebular line ratios using the same diagnostics applied to observational data over the full metallicity range covered by the model grid. We calculated these temperatures with \textsc{PyNeb} v1.1.30 \citep{Luridiana2015}, adopting the same atomic data as used by \textsc{Cloudy} C23.01 \citep[the Stout database;][]{Lykins2015,CloudyC23}. Consequently, any differences between the two temperature definitions arise from the thermal and ionization structure of the models rather than from inconsistencies in the adopted atomic data. The sensitivity of the derived temperatures to the adopted atomic data is discussed in Appendix~\ref{app:atomic}.

\subsection{Ion-weighted and line-derived temperatures}

To derive the line-based temperatures, we extracted the integrated emission-line intensities predicted by each \textsc{Cloudy} model, corresponding to the spectrum that would be observed from the entire nebula. Electron temperatures were then obtained from the integrated auroral-to-nebular line ratios using \textsc{PyNeb}. The resulting temperatures therefore reflect the integrated thermal and ionization structure of the nebula. The ion-weighted temperatures are the \textsc{Cloudy} averages of $T_{\rm e}$ over the nebular volume weighted by the electron and ion densities.
We then converted the [\ion{O}{III}]$\lambda\lambda$4959,5007/[\ion{O}{III}]$\lambda$4363,
[\ion{O}{II}]$\lambda\lambda$3727,3729/[\ion{O}{II}]$\lambda\lambda$7320,7330,
[\ion{N}{II}]$\lambda\lambda$6548,6584/[\ion{N}{II}]$\lambda$5755, and
[\ion{S}{III}]$\lambda\lambda$9069,9532/[\ion{S}{III}]$\lambda$6312 ratios into
$T_{\mathrm e}(\mathrm{[O\,III]})$,
$T_{\mathrm e}(\mathrm{[O\,II]})$,
$T_{\mathrm e}(\mathrm{[N\,II]})$, and
$T_{\mathrm e}(\mathrm{[S\,III]})$.
As emphasized by \citet{Garnett1992} and revisited by \citet{MendezDelgado2023} and \citet{RickardsVaught2024}, these line-derived temperatures may differ from the ion-weighted temperatures because auroral-to-nebular ratios give extra weight to the hottest and densest gas along the line of sight. They are therefore sensitive to local and global temperature variations within each ionization zone.

Fig.~\ref{figure:Te-comparison-young} shows the relative differences between the line-derived and ion-weighted electron temperatures for the 1--3~Myr models as a function of the ion-weighted temperature. $T_{\mathrm e}(\mathrm{[O\,II]})$ follows $T_{\mathrm e}(\mathrm{O}^{+})$ reasonably well, although larger systematic differences appear at both ends of the temperature range and show a dependence on the ionization parameter. At fixed $T_{\mathrm e}(\mathrm{O}^{+})$, $T_{\mathrm e}(\mathrm{[O\,II]})$ increases with $U$, so low-$U$ models cluster around and below the zero difference line, whereas high-$U$ models tend to lie above it. The relative difference stays within $\pm 4\%$ for low-$U$ models and reaches up to $10\%$ for high-$U$ models with very low $T_{\mathrm e}(\mathrm{O}^{+})$. For the vast majority of models, however, the relative difference remains within 5\%.

A similarly clear dependence on the ionization parameter is found for $T_{\mathrm e}(\mathrm{[N\,II]})$. At low $U$, $T_{\mathrm e}(\mathrm{[N\,II]})$ closely follows $T_{\mathrm e}(\mathrm{O}^{+})$, with typical differences of less than 2\%. At high $U$, however, a systematic trend is present: $T_{\mathrm e}(\mathrm{[N\,II]})$ exceeds $T_{\mathrm e}(\mathrm{O}^{+})$ by up to $\sim10\%$ at the lowest $T_{\mathrm e}(\mathrm{O}^{+})$, while the offset decreases steadily with increasing temperature, crossing the zero-difference line and reaching approximately $-3\%$ at the highest $T_{\mathrm e}(\mathrm{O}^{+})$.

A similarly good consistency, with a scatter of about 2\% around the zero line, is obtained for the high-ionization zone, where $T_{\mathrm e}(\mathrm{[O\,III]})$ closely follows $T_{\mathrm e}(\mathrm{O}^{++})$ over most of the grid. Only at the lowest temperatures, $T_{\mathrm e}(\mathrm{O}^{++}) \lesssim 7000$~K, $T_{\mathrm e}(\mathrm{[O\,III]})$ deviate appreciably, becoming higher than the ion-weighted value. This deviation is strongly dependent on ionization as it remains within $\sim 2\%$ at $\log U \sim -3$, but reaches a median of $\sim 15\%$ at $\log U \sim -1.5$ and $T_{\mathrm e}(\mathrm{O}^{++}) \sim 5000$~K.
It should be noted, however, that this region of parameter space corresponds to a combination of high metallicity (low electron temperature) and very high ionization parameter that is rarely occupied by observed nearby H\,{\sc ii} regions. $T_{\mathrm e}$ studies of metal-rich H\,{\sc ii} regions typically find $\log U \approx -3$, with an intrinsic scatter of only a few tenths of a dex \citep[e.g.,][]{PerezMontero2014,Sanchez2015,Zinchenko2019b,Garner2025}. The largest temperature offsets therefore arise mainly in an extreme corner of the model grid that is useful for exploring the underlying physics, but is expected to have limited impact on the vast majority of observed metal-rich nebulae.

The [\ion{S}{III}] diagnostic probes the S$^{++}$ zone and relies on [\ion{S}{III}]$\lambda$6312, a moderately bright auroral line detectable over most of the model grid. It exhibits a behavior similar to that of [\ion{N}{II}] (Fig.~\ref{figure:Te-comparison-young}). Since the S$^{++}$ zone overlaps substantially with the O$^{++}$ zone, $T_{\mathrm e}(\mathrm{[S\,III]})$ closely follows $T_{\mathrm e}(\mathrm{O}^{++})$ over most of the model grid, confirming that the S$^{++}$ zone temperature is a useful proxy for the O$^{++}$ zone temperature. At low $U$, the agreement is typically better than 3\%. At high $U$, a systematic trend is present: $T_{\mathrm e}(\mathrm{[S\,III]})$ exceeds $T_{\mathrm e}(\mathrm{O}^{++})$ by up to $\sim10\%$ at the lowest $T_{\mathrm e}(\mathrm{O}^{++})$, while the offset decreases with increasing temperature, reaching about $-6\%$ at the highest $T_{\mathrm e}(\mathrm{O}^{++})$. Compared to the [\ion{N}{II}] diagnostic, however, the transition is more gradual, with many models at intermediate temperatures showing only small offsets.

Figure~\ref{figure:Te-comparison-old} extends the comparison to the full 1--6~Myr grid in the age of the ionizing stellar population. The Wolf--Rayet (WR) contribution peaks at intermediate ages of $\sim 3$--$5$~Myr and hardens the ionizing radiation field at energies $\gtrsim 35$~eV. By 6~Myr, the most massive stars have died and the radiation field softens again. Figure~\ref{figure:Te-comparison-old} shows that the low-ionization $T_{\mathrm e}([\ion{O}{II}])$ and $T_{\mathrm e}([\ion{N}{II}])$ panels change little, although their high-$U$ branches become more dispersed. 
The scatter in the $T_{\mathrm e}([\ion{S}{III}])$--$T_{\mathrm e}(\mathrm{O}^{++})$ comparison at high $U$ also increases. Since the WR phase depends on metallicity, we examined how the extra scatter varies with O/H. It increases from 1.6\% to 5.0\% (1--3~Myr) and 3.1\% to 10.4\% (4--6~Myr) between $12+\log(\mathrm{O/H})=6.8$--7.4 and 8.6--9.2, with the older group remaining about twice as scattered.

Quantitatively, the fractions of models for which the line-derived temperatures deviate from the corresponding ion-weighted values by more than 5\% are 1.3\%, 1.5\%, 2.3\%, and 5.6\% for $T_{\mathrm e}([\ion{O}{II}])$, $T_{\mathrm e}([\ion{N}{II}])$, $T_{\mathrm e}([\ion{O}{III}])$, and $T_{\mathrm e}([\ion{S}{III}])$, respectively, over the full 1--6~Myr grid. These outliers are concentrated in the coldest, highest-$U$ corner of the grid. At $\log U \lesssim -3$, all four diagnostics agree to better than 5\% in essentially every model. The same plot, organized by the observable $\log(\mathrm{O32})$ rather than $\log U$, is shown in Appendix~\ref{app:logO32-residuals}.

\subsection{One-dimensional $t_2$--$t_3$ relation}
\label{sect:relations}

To parameterize the temperature relation, we first neglect its dependence on $U$ and fit $T_{\rm e}(\mathrm{O}^{+})$ as a quadratic function of $T_{\rm e}(\mathrm{O}^{++})$. We define
\begin{equation}
t_3 = \frac{T_{\mathrm e}(\mathrm{O}^{++})}{10^4~\mathrm{K}},
\qquad
t_2 = \frac{T_{\mathrm e}(\mathrm{O}^{+})}{10^4~\mathrm{K}},
\end{equation}
and fit the normalized temperature as
\begin{equation}
t_2 = a_1 + a_2 t_3 + a_3 t_3^2,
\label{eq:t2t3-fit-1D}
\end{equation}
independently for the Cloudy models with the age of ionizing source of 1--3, 4--6 and 1--6~Myr. The fit was obtained using an ordinary least-squares regression in the normalized variables $t_2$ and $t_3$, with all converged models weighted equally and no residual clipping applied. A quadratic form was adopted as the lowest-order polynomial capable of reproducing the curvature present in the model relation. Higher-order expansions did not significantly reduce the residual scatter and were therefore not considered further. 

\begin{figure}
\resizebox{1.00\hsize}{!}{\includegraphics[angle=000]{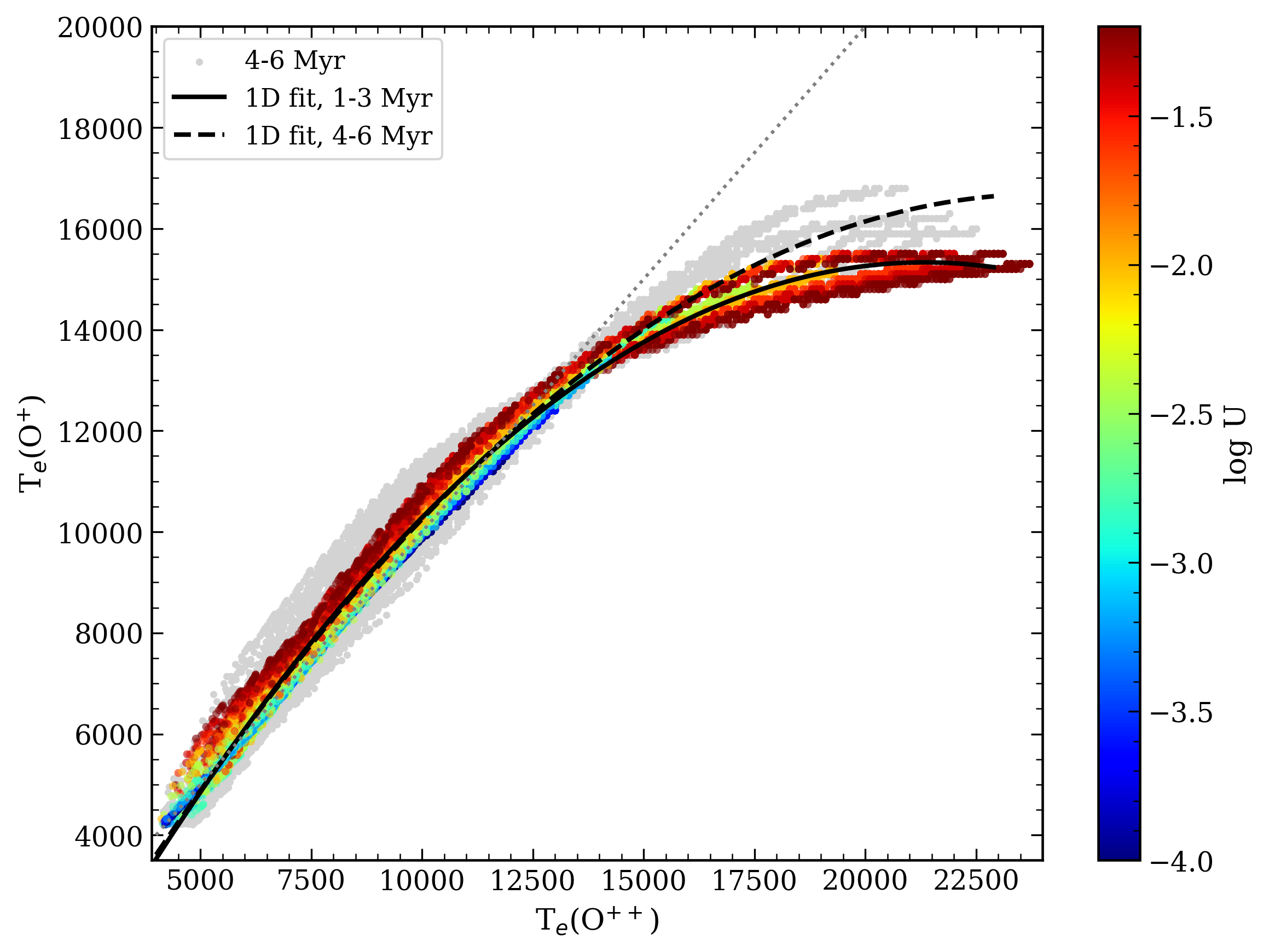}}
\caption{$T_{\mathrm e}(\mathrm{O}^+)$ versus $T_{\mathrm e}(\mathrm{O}^{++})$ for the 1--3~Myr models, color-coded by $\log U$. Gray circles are the 4--6~Myr models. The solid and dashed black lines are the one-dimensional quadratic fits for the 1--3 and 4--6~Myr age groups, respectively, and the dotted line is the one-to-one relation.}
\label{figure:t2t3}
\end{figure}

The relations for 1--3 and 4--6~Myr model grids are shown in Figure~\ref{figure:t2t3} by the black solid and dashed lines, respectively. In the low and intermediate temperature regime, up to $\sim 14\,000$~K, both relations show good consistency, although models ionized by older stellar populations show larger scatter around the fitting relation. For higher $t_3$ temperatures, $t_2$ for models ionized by older stellar populations starts to increase with $t_3$ faster than $t_2$ for models with younger stellar populations. For the highest $t_3$ the difference in $t_2$ reaches $\sim 1\,400$~K or 9\%. The robust scatter around the fit is 2.8\% and 4.3\% for the younger and older models, respectively. Also, models with older stellar populations show larger fraction of outliers with residuals exceeding 5\% for $T_{\mathrm e}(\mathrm{O}^{++}) \lesssim 12\,000$ K. 

\subsection{Second-parameter dependence}
\label{sect:second-param}

The one-dimensional relation presented above leaves a residual scatter of a few percent, which increases for the older ionizing populations. This scatter is not purely random since at fixed $T_{\mathrm e}(\mathrm{O}^{++})$, the models tend to order by ionization parameter, as shown in Fig.~\ref{figure:t2t3}, where the $t_2$--$t_3$ relation is color-coded by $\log U$ for the 1–3 Myr models. In this diagram, the high-$U$ models at low and intermediate temperature define the upper envelope of the relation. It is also clear that at high temperatures, the one-dimensional quadratic fits for the age groups 1--3 and 4--6~Myr deviate significantly. Therefore, a complete description of the temperature relation may require more than one additional parameter. 

Bearing that in mind, let us first consider a simplified picture. The ionization parameter largely determines the relative volumes occupied by the O$^{+}$ and O$^{++}$ ionization zones within an H\,{\sc ii} region.
At fixed $T_{\mathrm e}(\mathrm{O}^{++})$, increasing $U$ expands the O$^{++}$ zone and displaces the O$^{+}$ zone to larger depths within the nebula. Consequently, the gas emitting [\ion{O}{II}] lines is displaced to larger optical depths, where preferential absorption of low-energy ionizing photons progressively hardens the local radiation field. The resulting increase in the mean excess energy deposited by photoionization, together with the accompanying changes in the ionization balance, modifies the thermal balance of the O$^{+}$ zone, such that $T_{\mathrm e}(\mathrm{O}^{+})$ depends also on the ionization structure of the nebula. As a result, $T_{\mathrm e}(\mathrm{O}^{+})$ can deviate systematically from the value predicted by a simple one-parameter relation based solely on $T_{\mathrm e}(\mathrm{O}^{++})$. 

Since $U$ is not directly observable and is usually inferred from ionization-sensitive line ratios, we adopt the two ionization-sensitive ratios O32 and S2S3:
\begin{equation}
\mathrm{O32}=\frac{I([\ion{O}{iii}]\lambda5007)}{I([\ion{O}{ii}]\lambda\lambda3726,3729)},\quad
\mathrm{S2S3}=\frac{I([\ion{S}{ii}]\lambda\lambda6716,6731)}{I([\ion{S}{iii}]\lambda\lambda9069,9532)}.
\label{eq:ratios}
\end{equation}

The O32 ratio is a classical optical tracer of the nebular ionization state \citep{Kewley2002,Nakajima2014}, reflecting the relative abundance of O$^{++}$ and O$^{+}$. Although primarily sensitive to the ionization parameter, O32 also depends on the hardness of the ionizing spectrum and, more weakly, on metallicity \citep{PerezMontero2009,Dors2017}. Together with [\ion{S}{iii}]/[\ion{S}{ii}], it probes the ionization structure while responding to both $U$ and the shape of the ionizing SED, which is the basis of the softness diagram \citep{PerezMontero2009}. We also consider the \ion{He}{II}~$\lambda4686$/H$\beta$ ratio, which traces the hardness of the ionizing spectrum, and the electron density $n_{\mathrm e}$. Each tracer was used as the second variable in the two-dimensional fit and evaluated using 5-fold cross-validated root mean square error (RMSE) on a common sample of models.

\begin{figure*}
\resizebox{1.00\hsize}{!}{\includegraphics[angle=000]{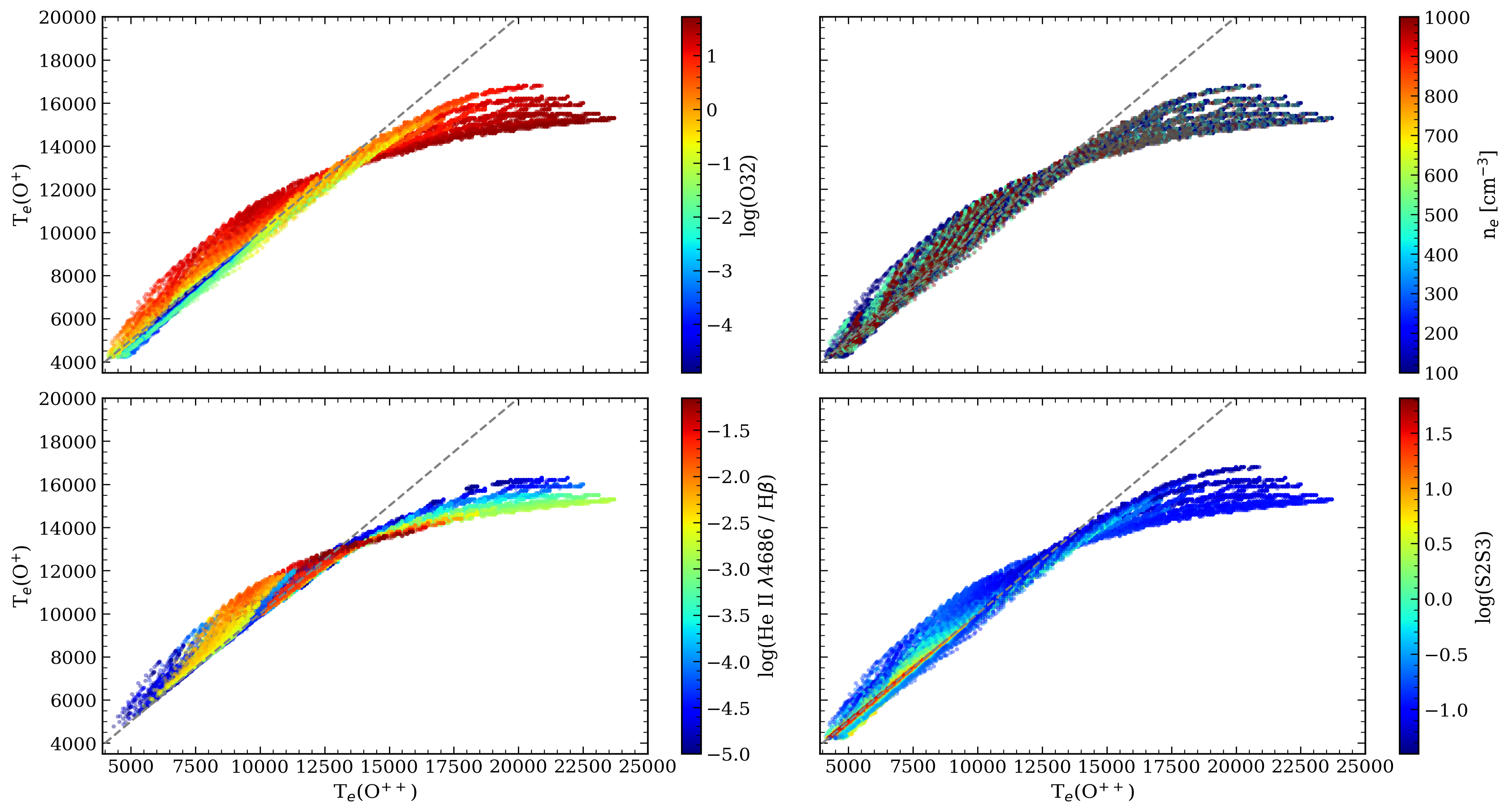}}
\caption{$T_{\rm e}({\rm O}^{+})$--$T_{\rm e}({\rm O}^{++})$ diagram for the 1--6~Myr grid color-coded by $\log(\mathrm{O32})$, electron density, $\log(\ion{He}{ii}\lambda4686/{\rm H}\beta)$, and $\log(\mathrm{S2S3})$. The dashed line is the one-to-one relation.
}
\label{figure:t2t3-colour}
\end{figure*}

Figure~\ref{figure:t2t3-colour} qualitatively illustrates the dependence of the $t_2$--$t_3$ relation on these secondary parameters. Color-coding the models by $\log(\mathrm{O32})$, $\log(\mathrm{S2S3})$, \ion{He}{II}/H$\beta$, and electron density demonstrates that a significant fraction of the scatter is caused by the ionization structure of the nebula and by the hardness of the ionizing radiation field, whereas no clear trend is apparent with electron density.

\begin{figure*}
\resizebox{1.00\hsize}{!}{\includegraphics[angle=000]{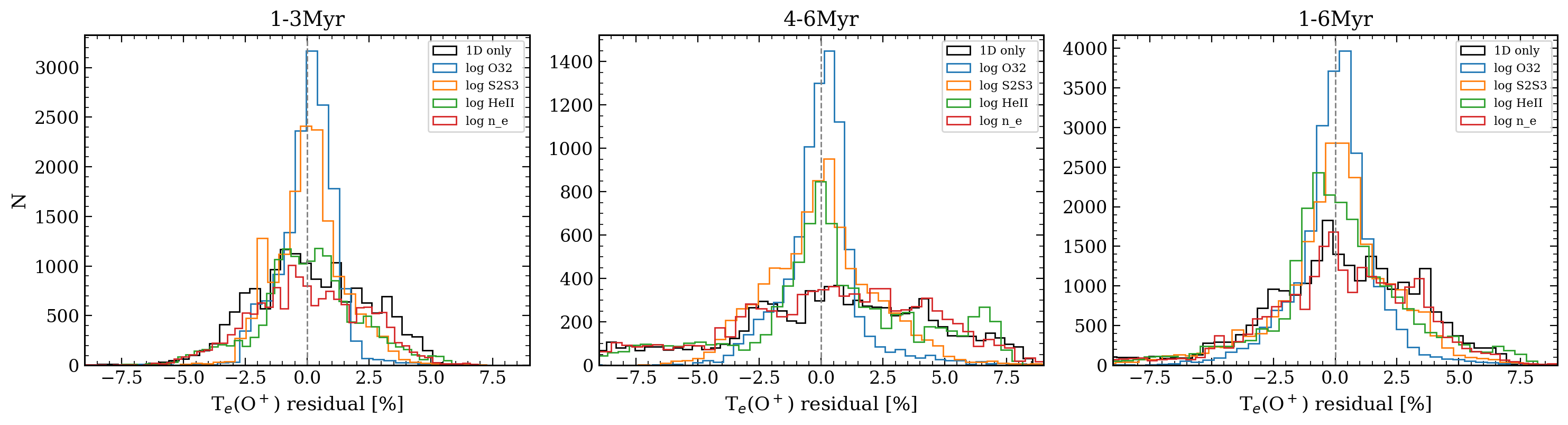}}
\caption{Residuals in $T_{\rm e}({\rm O}^{+})$ for the three age groups (1--3, 4--6, and 1--6~Myr). In each panel the black histogram is the one-dimensional fit (Eq.~\ref{eq:t2t3-fit-1D}) and the colored histograms are two-dimensional fits (Eq.~\ref{eq:t2t3-fit-2D}) using $\log(\mathrm{O32})$, $\log(\mathrm{S2S3})$, $\log(\ion{He}{ii}/{\rm H}\beta)$ or $\log n_{\rm e}$ as second parameter. The dashed line marks zero residual.
}
\label{figure:second-param-residuals}
\end{figure*}

Quantitatively, both $\log(\mathrm{O32})$ and $\log(\mathrm{S2S3})$ reduce the 5-fold cross-validated RMSE of the one-dimensional relation by 40--60\%, with nearly identical performance across the three stellar-age bins. The corresponding residual distributions for the three age groups are shown in Fig.~\ref{figure:second-param-residuals}. The \ion{He}{II}~$\lambda4686$/H$\beta$ ratio is less effective overall and is undefined for a substantial fraction of the model grid. Nevertheless, it exhibits a clear correlation with $T_{\mathrm e}(\mathrm{O}^{+})$ at high temperatures, $T_{\mathrm e}(\mathrm{O}^{++}) \gtrsim 14\,000$~K, where models with stronger \ion{He}{II} emission systematically show lower $T_{\mathrm e}(\mathrm{O}^{+})$ at fixed $T_{\mathrm e}(\mathrm{O}^{++})$. By contrast, the electron density provides little additional predictive power. 

In all three age groups, the inclusion of $\log(\mathrm{O32})$ or $\log(\mathrm{S2S3})$ substantially narrows the residual distribution and largely removes the extended wings present in the one-dimensional fits, particularly for the 4--6~Myr models. We therefore adopt $\mathrm{O32}$ as the preferred secondary parameter, as it performs slightly better than $\mathrm{S2S3}$. We incorporate this dependence by extending the $t_2$--$t_3$ relation to a two-dimensional polynomial:
\begin{equation}
t_2 = c_1 + c_2 t_3 + c_3 o + c_4 t_3\,o +
c_5 t_3^2 + c_6 o^2 + c_7 t_3^2 o^2 + c_8 t_3^3,
\label{eq:t2t3-fit-2D}
\end{equation}
where 
\begin{equation}
o = \log(\mathrm{O32}).
\end{equation}
The same form with $s=\log(\mathrm{S2S3})$ in place of $o$, fitted to the same age groups, serves spectra without [\ion{O}{II}]$\lambda\lambda$3726,3729. Its coefficients $b_1$--$b_8$ are given in the lower block of Table~\ref{tab:coeffs}.

The relation was fitted separately to the same three stellar-age groups used for the one-dimensional calibration. The coefficients of both the one- and two-dimensional fits, together with the corresponding scatters, are listed in Table~\ref{tab:coeffs}.
For the two-dimensional fit, the largest residuals, both positive and negative, are confined to the low temperature high-$O32$ corner of the parameter space as well as the lower edge of $t_3$ range.

\begin{table*}
\caption{Coefficients of the $t_2$--$t_3$ relations derived in this work.}
\label{tab:coeffs}
\centering
\small
\setlength{\tabcolsep}{3pt}
\begin{tabular}{lcccc}
\hline\hline
\multicolumn{5}{c}{One-dimensional fit:\quad $t_2 = a_1 + a_2\,t_3 + a_3\,t_3^2$}\\
\hline
Age group & $a_1$ & $a_2$ & $a_3$ & scatter \\
\hline
1--3~Myr & $-0.254779$ & $1.675586$ & $-0.392589$ & $2.8\%$ \\
4--6~Myr & $-0.210843$ & $1.559133$ & $-0.323324$ & $4.3\%$ \\
1--6~Myr & $-0.256102$ & $1.663569$ & $-0.378678$ & $3.7\%$ \\
\hline
\end{tabular}

\begin{tabular}{lccccccccc}
\hline\hline
\multicolumn{10}{c}{Two-dimensional fit (Eq.~\ref{eq:t2t3-fit-2D}), with $o=\log(\mathrm{O32})$}\\
\hline
Age group & $c_1$ & $c_2$ & $c_3$ & $c_4$ & $c_5$ & $c_6$ & $c_7$ & $c_8$ & scatter \\
\hline
1--3~Myr & $-0.160289$ & $1.512694$ & $0.084150$ & $-0.053551$ & $-0.355613$ & $0.024659$ & $-0.016277$ & $0.024844$ & $1.2\%$ \\
4--6~Myr & $-0.024836$ & $1.271876$ & $0.144812$ & $-0.104216$ & $-0.235446$ & $0.021588$ & $-0.016116$ & $0.022125$ & $1.9\%$ \\
1--6~Myr & $-0.029045$ & $1.228681$ & $0.121334$ & $-0.087291$ & $-0.163750$ & $0.019887$ & $-0.015787$ & $-0.008213$ & $1.8\%$ \\
\hline
\end{tabular}

\begin{tabular}{lccccccccc}
\hline\hline
\multicolumn{10}{c}{Two-dimensional fit (Eq.~\ref{eq:t2t3-fit-2D}) with $s=\log(\mathrm{S2S3})$ in place of $o$}\\
\hline
Age group & $b_1$ & $b_2$ & $b_3$ & $b_4$ & $b_5$ & $b_6$ & $b_7$ & $b_8$ & scatter \\
\hline
1--3~Myr & $-0.201333$ & $1.489608$ & $-0.022264$ & $-0.004162$ & $-0.248253$ & $0.035756$ & $-0.021406$ & $-0.029974$ & $1.5\%$ \\
4--6~Myr & $-0.075183$ & $1.181408$ & $-0.087487$ &  $0.041889$ & $-0.023240$ & $0.033141$ & $-0.003891$ & $-0.077401$ & $2.3\%$ \\
1--6~Myr & $-0.103604$ & $1.243736$ & $-0.053213$ &  $0.019429$ & $-0.036885$ & $0.005972$ & $0.008075$  & $-0.091836$ & $2.1\%$ \\
\hline
\end{tabular}
\tablefoot{
The upper part lists the coefficients of the one-dimensional quadratic fit. The middle part gives the coefficients of the two-dimensional fit using $o=\log(\mathrm{O32})$, while the lower part gives the corresponding fit using $s=\log(\mathrm{S2S3})$. The last column gives the robust scatter of the predicted $T_{\rm e}(\mathrm{O}^{+})$. The relations derived for the 1--6~Myr models are recommended for samples with mixed or unknown ionizing-population ages. Sampling uncertainties of the polynomial coefficients are negligible compared to the model systematic uncertainties.
}

\end{table*}

\subsection{Dependence on the ionizing SED}
\label{sect:sed}

The POPSTAR SEDs used so far are based on single-star evolution. Binary evolution can harden the ionizing spectrum, especially at ages beyond a few Myr, and may therefore alter the $t_2$--$t_3$ relation. To assess this effect, we recomputed the model grid using BPASS v2.3 \citep{Eldridge2017,BPASS23_p2,Byrne2022} ionizing SEDs, adopting both the binary and single-star prescriptions, and compared the results with the POPSTAR grid over the full 1--6~Myr age range.

\begin{figure*}
\resizebox{1.00\hsize}{!}{\includegraphics[angle=000]{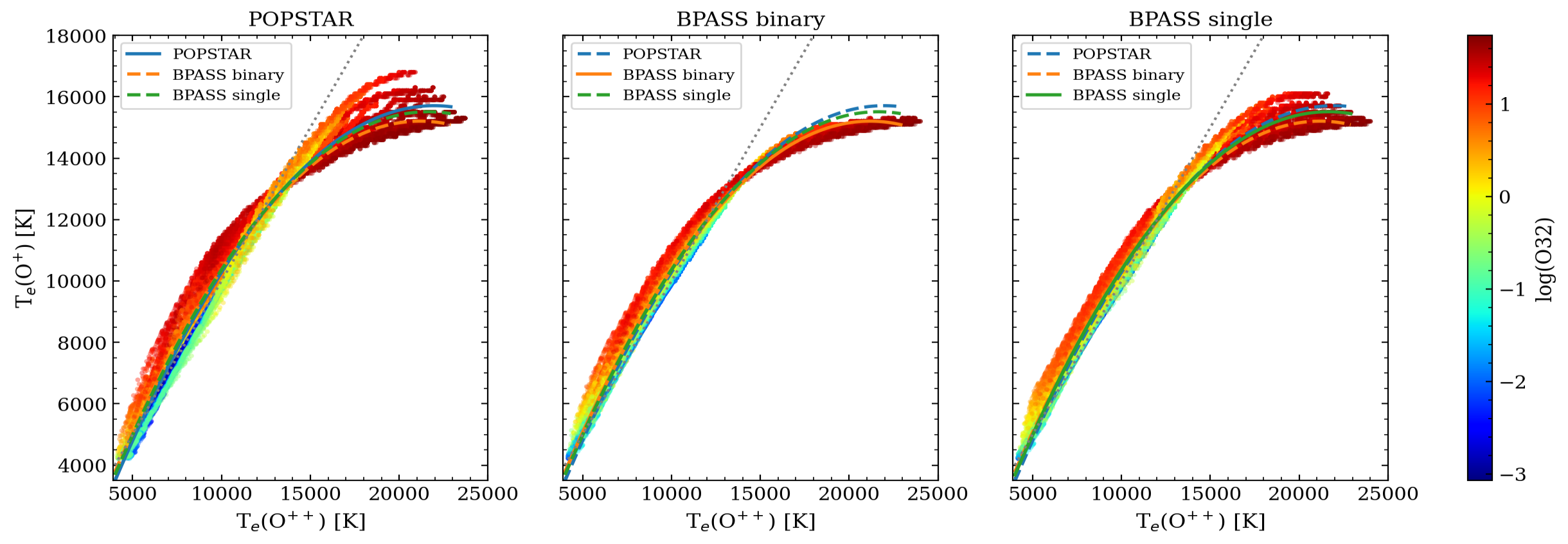}}
\caption{$t_2$--$t_3$ relation for POPSTAR, BPASS binary, and BPASS single-star ionizing SEDs over the age range 1--6~Myr. Lines show the one-dimensional quadratic fits for each SED family. The dotted line marks the one-to-one relation.}
\label{figure:sed}
\end{figure*}

Figure~\ref{figure:sed} shows the resulting $t_2$--$t_3$ relations. The three SED families follow very similar trends: the one-dimensional fits agree to within a few percent in $t_2$ across the full temperature range. The scatter, however, depends on the adopted SED family. The one-dimensional fits yield RMSE values of 427, 281, and 339~K for the POPSTAR, BPASS binary, and BPASS single-star grids, respectively, while the two-dimensional fits reduce these values to 248, 133, and 188~K. The slightly larger scatter in the POPSTAR grid likely reflects its more concentrated WR phase at $3$--$5$~Myr, whereas the binary stripped stars in BPASS harden the radiation field more gradually with age. These differences are small compared with the dominant ionization-parameter dependence quantified in Sect.~\ref{sect:second-param}. For the POPSTAR and BPASS configurations considered here, the ionizing SED is therefore a secondary effect. Since the comparison simultaneously changes stellar evolution, the WR prescription, binarity, and the IMF, this conclusion should not be generalized to all population synthesis models.

\subsection{The effect of dust}
\label{sect:dust}

\begin{figure}
\resizebox{1.00\hsize}{!}{\includegraphics[angle=000]{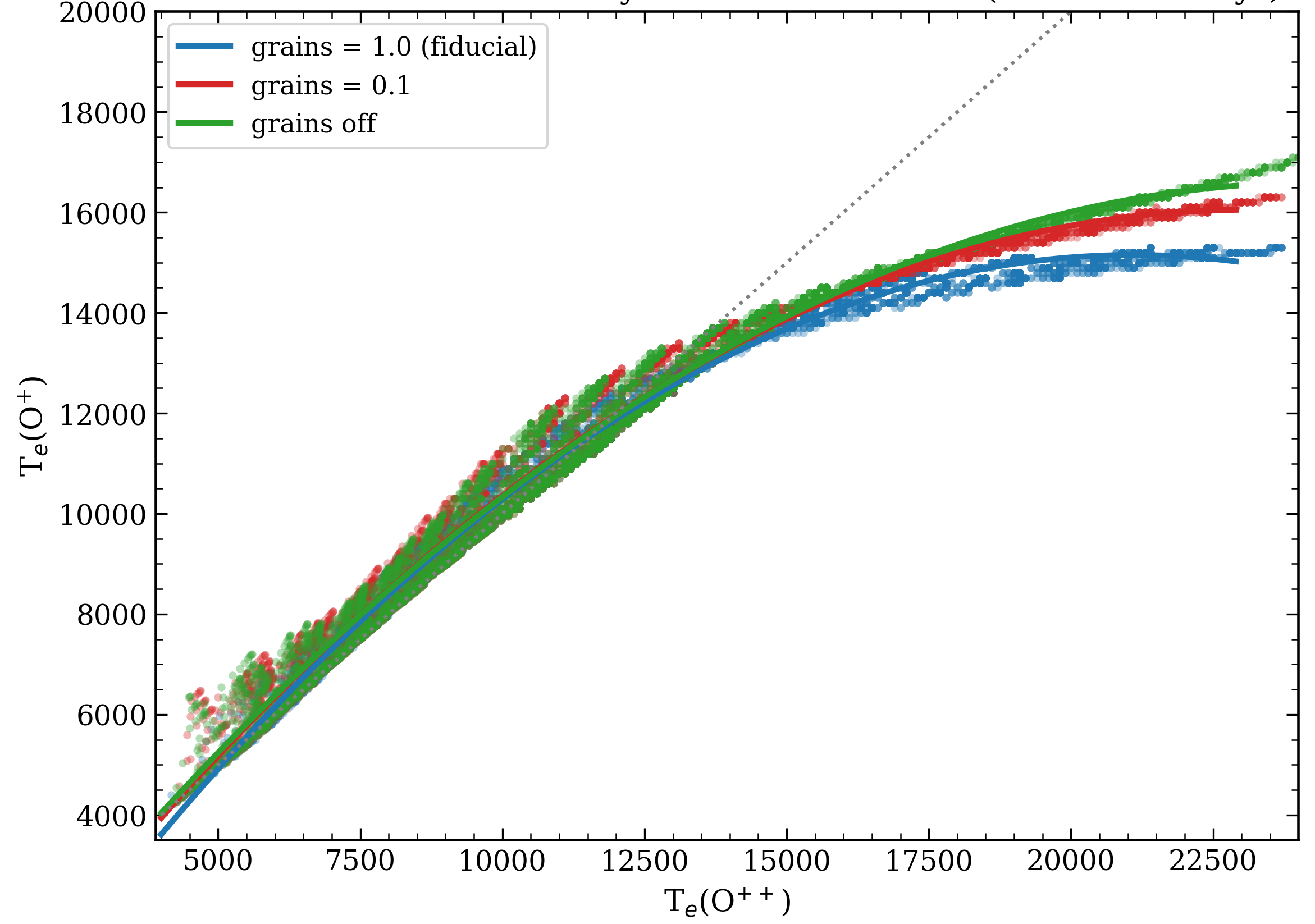}}
\caption{Effect of the dust grain abundance on the $t_2$--$t_3$ relation, for the POPSTAR 1~Myr grid. Points and curves show the $T_{\rm e}(\mathrm{O}^{+})$--$T_{\rm e}(\mathrm{O}^{++})$ models and the one-dimensional quadratic fit for the fiducial grain abundance (grains~$=1.0$, blue), the reduced-grain variant (grains~$=0.1$, red), and the grains-off variant (green). The dotted line marks the one-to-one relation.}
\label{figure:grains}
\end{figure}

Dust is another modeling component capable of producing substantial systematic changes in the predicted temperature structure. To isolate its influence, we compare the fiducial grid, which adopts the standard \textsc{Cloudy} ISM grain mixture ($\mathrm{grains}=1$, corresponding to the Milky Way dust-to-gas mass ratio), with two otherwise identical model sets computed at 1~Myr: one with the grain abundance reduced by an order of magnitude ($\mathrm{grains}=0.1$) and one with the grain component removed entirely. Figure~\ref{figure:grains} illustrates how the dust proportion affects the $t_2$--$t_3$ relation.
The response to changing the dust content is notably differential. Reducing the grain abundance substantially raises $T_{\rm e}({\rm O}^{+})$ at high temperatures, whereas $T_{\rm e}({\rm O}^{++})$ remains nearly unchanged. The effect decreases steadily toward lower $T_{\rm e}({\rm O}^{++})$ and becomes negligible for $T_{\rm e}({\rm O}^{++})\lesssim16\,000$~K.

Because the two ionization zones respond differently, changing the dust content modifies the slope rather than simply shifting the $t_2$--$t_3$ relation. Dust-poor models therefore predict systematically higher $t_2$ at fixed $t_3$, with the difference increasing toward higher temperatures and excitation. The ordering of the residuals by $O_{32}$ is largely preserved across all dust prescriptions. Although its predictive power decreases slightly as the grain abundance is reduced, $O_{32}$ remains the most effective secondary parameter. Dust thus introduces a systematic component that cannot be removed by accounting for the ionization structure alone.

Overall, these experiments identify dust as one of the dominant remaining sources of systematic uncertainty in the theoretical calibration at high temperatures. The form of the $t_2$--$t_3$ relation and the importance of the ionization structure remain unchanged, but the absolute calibration depends on the adopted dust prescription.

\section{Discussion}
\label{section:Discussion}

Our results indicate that the $t_2$--$t_3$ relation derived from a modern photoionization grid of \ion{H}{II} regions is neither universal nor strictly one-dimensional. Instead, it depends on the ionization parameter, traced observationally by the ionization-sensitive ratios O32 and S2S3, neither of which is determined by $U$ alone (Sect.~\ref{sect:second-param}), and on the hardness of the ionizing radiation field propagating through an \ion{H}{II} region. The relations derived here should therefore be regarded as model-calibrated parameterizations valid within the explored range of physical conditions, rather than universal empirical laws. In particular, they apply to the parameter space covered by our grid: $12+\log(\mathrm{O/H})=6.8$--9.2, $\log U=-4.0$ to $-1.2$, $\log(\mathrm{N/O})=-2.0$ to $0.0$, electron densities of 100--1000~cm$^{-3}$, and ionizing-source ages of 1--6~Myr. They should be used with caution for nebulae dominated by much harder radiation fields, such as AGN- or X-ray binary-ionized systems, for objects with strong shock contributions, or for systems with complex morphology and/or strongly clumpy density structures. 

\begin{figure*}
\resizebox{1.00\hsize}{!}{\includegraphics[angle=000]{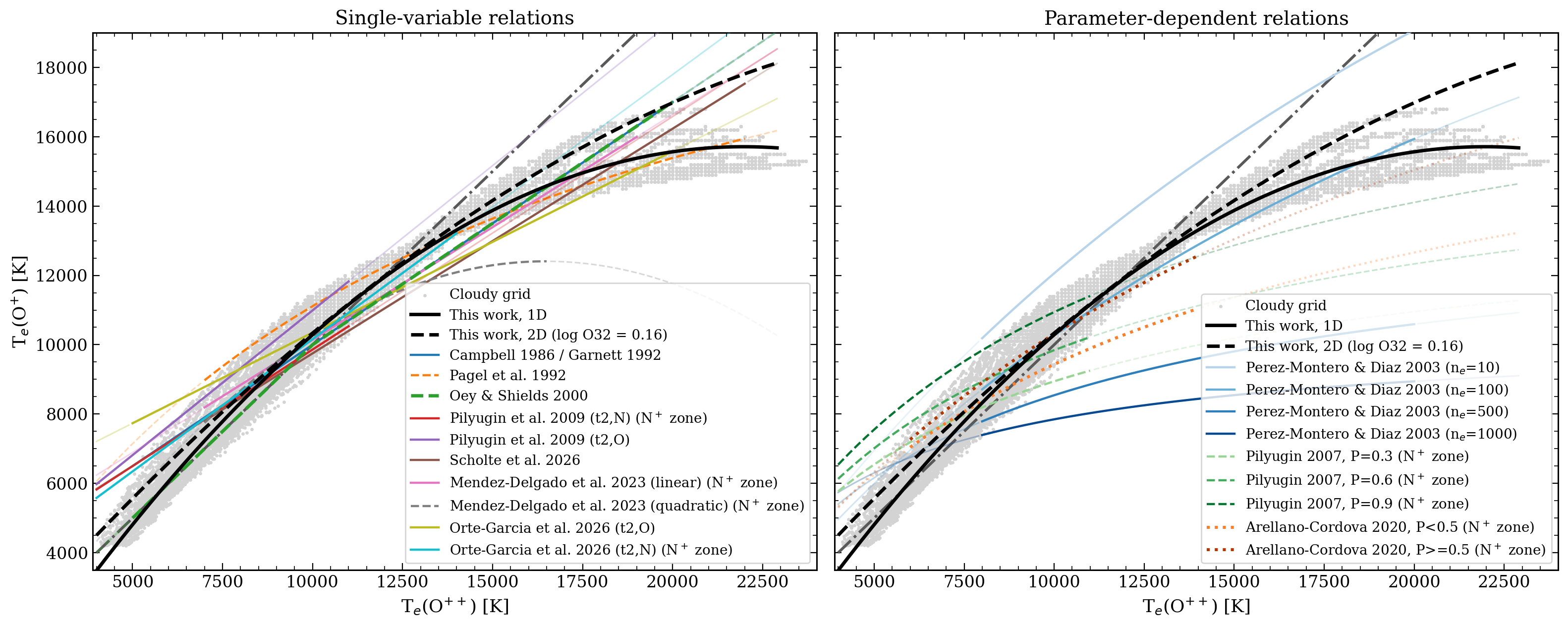}}
\caption{$T_{\rm e}(\mathrm{O}^{+})$--$T_{\rm e}(\mathrm{O}^{++})$ plane for the 1--6~Myr Cloudy grid (gray points), with the one- and two-dimensional relations of this work and a set of $t_2$--$t_3$ relations from the literature. The left panel shows single-variable relations, the right panel parameter-dependent ones. Each relation is drawn solid within its approximate validity range and thin outside it. The dash-dotted gray line is the one-to-one relation.
}
\label{figure:literature-comparison}
\end{figure*}

\subsection{Physical origin of the nonlinearity and scatter}
\label{sect:physical-origin}

The electron temperature of a photoionized nebula is determined by the balance between radiative heating and cooling. Heating is dominated by photoelectrons produced by hydrogen and helium ionization, whereas cooling is generally dominated by collisionally excited metal lines, although hydrogen and helium become increasingly important at very low metallicities. Because metallicity influences the thermal balance primarily through its effect on the overall cooling efficiency, much of its impact is already encoded in the equilibrium electron temperature itself. Consistent with this expectation, once $T_{\rm e}({\rm O}^{++})$ is specified, the residual structure of our model grid is organized predominantly by ionization parameter rather than by metallicity.

The nonlinearity of the $t_2$--$t_3$ relation, the apparent saturation of $t_2$ at high $t_3$, and the stratification of the models by parameters connected with $U$ can therefore be understood as different manifestations of the changing thermal and geometric relationship between the O$^{+}$ and O$^{++}$ zones. At high ionization parameter the O$^{+}$ region becomes a relatively thin transition layer near the outer edge of the \ion{H}{II} region, causing the temperatures of the two zones to respond differently to increasing excitation. The resulting departure from a one-to-one mapping produces both the observed nonlinearity and the strong dependence on O32 and S2S3.

Previous studies have attributed departures from a one-dimensional $t_2$--$t_3$ sequence to excitation effects \citep{Pilyugin2007,Pilyugin2009,Pilyugin2009SDSS}, density dependence \citep{PerezMontero2003}, or internal temperature structure \citep{MendezDelgado2023,RickardsVaught2024}. Our models suggest that these apparently distinct phenomena can be explained within a framework in which the ionization parameter alters the relative geometry and thermal balance of the O$^{+}$ and O$^{++}$ zones. This interpretation rests, however, on a deliberately idealized model whose limitations should be taken into account when applying the proposed relations. The calculations assume plane-parallel geometry and constant density, whereas real \ion{H}{II} regions exhibit complex density structure and turbulence. Also, mechanical heating from shocks and stellar feedback is neglected and all models assume a Maxwell electron energy distribution.

The adopted dust prescription is an additional source of uncertainty. As shown in Figure~\ref{figure:grains}, varying the dust abundance modifies the thermal balance of the nebula and shifts the fitted $t_2$--$t_3$ relation by several percent, comparable to the intrinsic scatter of the one-dimensional calibration. The present calibrations therefore implicitly assume dust properties similar to the standard \textsc{Cloudy} ISM grain model. Since both the dust-to-gas ratio and grain properties evolve with metallicity and galaxy environment \citep[e.g.,][]{RemyRuyer2014,Galliano2018,DeVis2019}, caution is needed when applying these relations to systems with dust properties that differ substantially from those assumed here. We plan to study this issue in detail in future work.

Finally, integrated galaxy spectra may combine emission from multiple unresolved \ion{H}{II} regions, so the effective temperature relation may not coincide with that of any individual nebula. The calibrations presented here apply most directly to individual \ion{H}{II} regions, and their use for spatially integrated galaxy spectra still requires observational validation, deferred to Paper~II. None of these idealizations alters the qualitative origin of the nonlinearity established above, but together they bound the quantitative accuracy of the calibration.

\begin{table*}
\caption{Comparison of published $t_2$--$t_3$ relations with the photoionization grid.}
\label{tab:literature}
\centering
\small
\begin{tabular}{llcccc}
\hline\hline
Reference & Type & Parameters & \multicolumn{3}{c}{RMSE [\%]} \\
\cline{4-6}
 & & & all & $t_3<1.4$ & $t_3>1.4$ \\
\hline
\citet{Pagel1992}           & theor., non-linear      & $t_3$       & 14.6 & 17.9 & 3.9 \\
\citet{Garnett1992}         & theor., linear      & $t_3$       & 9.8  & 10.5 & 8.2 \\
\citet{Oey2000}             & theor., broken      & $t_3$       & 6.5  & 5.4  & 8.2 \\
\citet{PerezMontero2003}    & theor., non-linear  & $t_3, n_e$  & 7.4  & 8.4  & 5.0 \\
\citet{PerezMontero2014}, O$^{+}$ zone & theor., non-linear & $t_3$ & 6.9 & 7.6 & 5.4 \\
\citet{PerezMontero2014}, N$^{+}$ zone$^\dagger$ & theor., non-linear & $t_3$ & 8.4 & 8.5 & 8.4 \\
\citet{Pilyugin2007}$^\dagger$ & empir., non-linear  & $t_3, P$    & 17.5 & 16.1 & 19.9 \\
\citet{Pilyugin2009}, [\ion{N}{II}]$^\dagger$ & empir., linear      & $t_3$       & 9.7  & 10.7 & 7.5 \\
\citet{Pilyugin2009}, [\ion{O}{II}]    & empir., linear      & $t_3$       & 16.3 & 15.1 & 18.4 \\
\citet{ArellanoCordova2020}$^\dagger$ & empir., non-linear  & $t_3, P$    & 9.4  & 10.7 & 6.5 \\
\citet{MendezDelgado2023}, linear$^\dagger$    & empir., linear      & $t_3$       & 11.3 & 12.9 & 7.2  \\
\citet{MendezDelgado2023}, quadratic$^\dagger$ & empir., non-linear  & $t_3$       & 12.7 & 7.5  & 18.9 \\
\citet{Scholte2026}        & empir., linear       & $t_3$       & 10.0 & 11.1 & 7.5 \\
\citet{OrteGarcia2026}, [\ion{O}{II}]   & empir., linear      & $t_3$       & 15.5 & 18.6 & 6.8 \\
\citet{OrteGarcia2026}, [\ion{N}{II}]$^\dagger$ & empir., linear      & $t_3$       & 10.0 & 9.6  & 10.9 \\
This work (1D)              & theor., quadratic   & $t_3$       & 4.2  & 4.7  & 3.0 \\
This work (2D)              & theor., polynomial  & $t_3,$ O32  & 2.5  & 2.8  & 1.9 \\
\hline
\end{tabular}
\tablefoot{
The last three columns list the root-mean-square error (RMSE) between the $t_2$ predicted by each relation and the ion-weighted $t_2$ of the full 1--6~Myr grid, computed for the entire $t_3$ range and separately for $t_3<1.4$ and $t_3>1.4$. Relations marked with $\dagger$ are calibrated using $T_{\rm e}(\mathrm{N}^{+})$ rather than $T_{\rm e}(\mathrm{O}^{+})$; consequently, the quoted RMSE also includes the systematic offset between the N$^{+}$ and O$^{+}$ temperature zones. The RMSE is calculated over the full distribution and therefore differs slightly from the MAD-based scatter quoted for our temperature relations in Table~\ref{tab:coeffs}.
}

\end{table*}

\begin{figure*}
\resizebox{1.00\hsize}{!}{\includegraphics[angle=000]{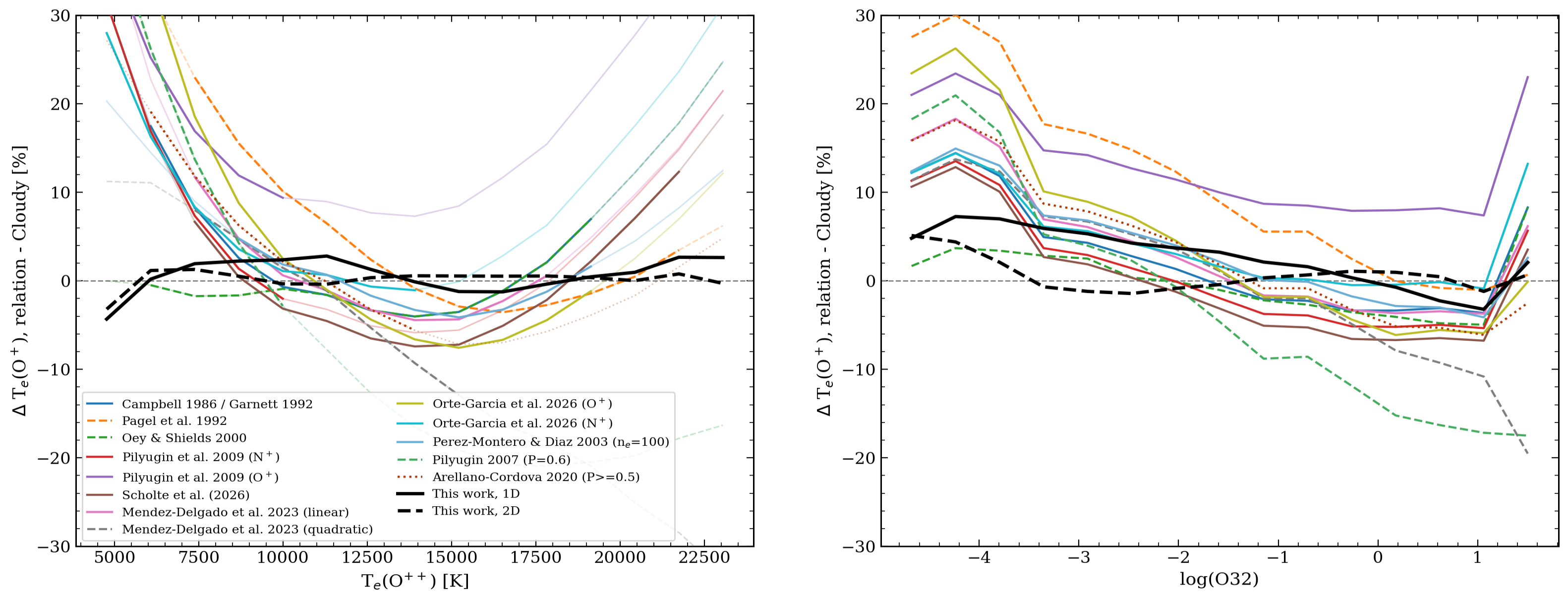}}
\caption{Residuals of literature $t_2$--$t_3$ relations with respect to the ion-weighted $t_2$ of the 1--6~Myr grid, as a function of $T_{\rm e}({\rm O}^{++})$ and $\log(\mathrm{O32})$.}
\label{figure:classical-residuals}
\end{figure*}

\subsection{Comparison with previous studies}
\label{sect:literature-comparison}

The classical theoretical $t_2$--$t_3$ relations \citep{Stasinska1980,Campbell1986,Pagel1992,Garnett1992} assumed that the temperatures of the O$^{+}$ and O$^{++}$ zones are linked by a one-to-one relation. The linear calibration of \citet{Garnett1992}, in particular, has become the standard prescription in direct-method abundance studies. Figure~\ref{figure:literature-comparison} compares a set of calibrations collected from the literature with our Cloudy model grid. Most literature relations reproduce the general trend of increasing $t_2$ with $t_3$, but none simultaneously matches the low- and high-ionization branches of the model sequence in the entire temperature range. The underlying nonlinearity is strongest at high $t_3$. As a result, the classical relations systematically over- or under-predict $t_2$ in different regions of parameter space.

These differences are quantified in Table~\ref{tab:literature}, which lists, for each published relation, the RMS difference from the ion-weighted $t_2$ of the full model grid for both full $t_3$ range and separately for the low- ($t_3<1.4$) and high-temperature ($t_3>1.4$) branches. Splitting the residuals at $T_{\rm e}(\mathrm{O}^{++})=14\,000$~K is informative because the grid is dominated by cooler, lower-ionization models, so the overall RMSE is weighted toward the lower $t_3$ and underestimates the scatter in the hot, highly ionized regime that is increasingly probed at high redshift and local metal-poor galaxies. The split shows that most classical calibrations reach substantial accuracy in only one of the two regimes. Strongly saturating relations such as that of \citet{Pagel1992} reproduce the high-$t_3$ flattening well but overestimate $t_2$ at low $t_3$, whereas the quadratic relation of \citet{MendezDelgado2023}, fitted to 42 DESIRED \ion{H}{ii} regions, is accurate at low $t_3$ but turns over and falls away at high $t_3$, where its shape above 13\,000~K is set by only two objects. The linear relation from the same study shows the opposite trade-off.
Among the one-dimensional literature calibrations the broken-linear relation of \citet{Oey2000} has the lowest overall RMSE. The split demonstrates that this advantage is caused by its low-$t_3$ isothermal segment, which matches the behavior of the grid, especially at low $U$, whereas its high-$t_3$ branch, identical to that of \citet{Garnett1992}, still misses part of the saturation. 

The behavior of the residuals is illustrated in Fig.~\ref{figure:classical-residuals}. The residuals of the $T_{\rm e}(\mathrm{O}^{+})$ computed by literature relations exhibit systematic trends with both $T_{\rm e}(\mathrm{O}^{++})$ and $\log({\rm O32})$. Most literature calibrations show their best agreement with the model grid at intermediate temperatures, close to the parameter range over which they were originally derived. At both low and high $T_{\rm e}(\mathrm{O}^{++})$, however, the deviations become progressively larger.

A second feature is the significant dependence of the residuals on $\log({\rm O32})$. At fixed $T_{\rm e}(\mathrm{O}^{++})$, models with different ionization states are systematically offset from one another, producing residual patterns rather than random scatter. The same behavior is visible across nearly all of the published calibrations. In this sense, the observed scatter in the classical $t_2$--$t_3$ relation is not simply measurement uncertainty or model dispersion, but reflects the projection of a multidimensional relation onto a single temperature coordinate.

Several previous studies anticipated aspects of this behavior. \citet{Oey2000} introduced a broken-linear parameterization that reproduces part of the flattening observed at high $t_3$, although without an explicit dependence on ionization state. Likewise, the non-linear relation proposed by \citet{PerezMontero2003} captures some of the curvature seen in the model grid, although it assumes a density dependence, which was obtained for line-derived rather than ionic temperatures, as shown later by \citet{PerezMontero2014}. In our models, however, electron density contributes only weakly to the residual scatter and cannot account for the systematic trends. This applies to a constant density between 100 and 1000~cm$^{-3}$ and does not test internal gradients or clumping, which may act differently. Detailed studies of density variations are needed to draw a more general conclusion.

The closest conceptual precedent is provided by \citet{Pilyugin2007,Pilyugin2009,Pilyugin2009SDSS}, who demonstrated empirically that high-metallicity \ion{H}{II} regions do not follow a unique $t_2$--$t_3$ sequence and introduced the excitation parameter $P$ as an additional coordinate. Because both $P$ and O32 are sensitive to the ionization state of the nebula, while retaining some dependence on the ionizing SED and on metallicity, our relation $t_2=f(t_3,\log{\rm O32})$ may be viewed as the theoretical counterpart of that empirical result. Recent observational studies are broadly consistent with this interpretation. The increased scatter and apparent saturation reported by \citet{ArellanoCordova2020} and \citet{MendezDelgado2023} are reproduced by our model grid and emerge naturally once the ionization parameter is allowed to vary. Likewise, integral-field analyses, metal-rich auroral-line studies, and large spectroscopic compilations \citep{RickardsVaught2024,Peng2026,OrteGarcia2026} consistently find relations based on $T_{\rm e}([\ion{N}{II}])$ to exhibit the smallest dispersion and the closest correspondence to temperatures in the low-ionization zone. This is in good agreement with our finding that $T_{\rm e}([\ion{N}{II}])$ provides a reliable observational tracer of $T_{\rm e}({\rm O}^{+})$. The shallower $T_{\rm e}([\ion{O}{II}])$--$T_{\rm e}([\ion{O}{III}])$ slope reported for high-redshift galaxies by \citet{Cataldi2025} is also expected if such systems preferentially occupy the high temperature and high O32 area of the two-dimensional relation. In this context, the shallower $T_{\rm e}([\ion{O}{II}])$--$T_{\rm e}([\ion{O}{III}])$ slopes reported for high-redshift galaxies may reflect the preferential sampling of high temperature and high-ionization environments rather than any fundamental change in the thermal physics of \ion{H}{II} regions. A quantitative comparison between the dispersion of the present grid and the scatter actually observed in these samples would establish whether the modeled effects account for the full observed dispersion. That test requires a homogeneous observational compilation with realistic uncertainties and is deferred to Paper~II.

\subsection{Line-derived versus ion-weighted temperatures}
\label{sect:obs-vs-phys}

Another important aspect of this work is the characterization of the systematic difference between ion-weighted temperatures and temperatures derived from auroral-to-nebular line ratios. The former describe the physical thermal structure of the nebula and are therefore the natural quantities predicted by photoionization models, whereas the latter are the quantities accessible to observations. Distinguishing between these two definitions is essential when comparing theoretical and empirical temperature relations.

Because the emissivities of auroral lines depend exponentially on temperature, the hottest regions within an ionization zone contribute disproportionately to the integrated auroral emission. Temperatures derived from auroral-to-nebular line ratios therefore tend to be weighted toward the warmest gas rather than representing the corresponding ion-weighted temperatures. In the present models, this offset arises from the continuous temperature and ionization structure predicted by photoionization theory, without invoking additional temperature fluctuations beyond those produced by the model itself.

For most of the parameter space explored here, the differences between line-derived and ion-weighted temperatures remain modest, typically within a few percent. However, the discrepancy increases in the coldest and most highly ionized models, reaching up to $\sim20\%$ for $T_{\rm e}([\ion{O}{III}])$. These extreme cases correspond to the same low temperature and high-$U$ regime in which the classical one-dimensional $t_2$--$t_3$ relations perform least well. The effect is therefore not merely a technical detail of temperature diagnostics but a potentially significant source of systematic uncertainty when interpreting highly ionized nebulae.

$T_{\rm e}([\ion{N}{II}])$ follows $T_{\rm e}({\rm O}^{+})$ as closely as $T_{\rm e}([\ion{O}{II}])$, with differences reaching up to $\sim10\%$ only at low temperature and high ionization parameter. In practice, the [\ion{O}{II}]$\lambda\lambda7320,7330$ diagnostic deserves particular caution because it is more sensitive to density and may receive a non-negligible recombination contribution. Together with recent observational studies, these results support the recommendation to use $T_{\rm e}([\ion{N}{II}])$ to trace temperature in the low ionization zone as a reliable substitute of $T_{\rm e}([\ion{O}{II}])$.

\subsection{Implications for abundance determinations}
\label{sect:abundance-implications}

In most \ion{H}{II} regions, only one of the two characteristic electron temperatures is measured directly, while the other is inferred from a $t_2$--$t_3$ relation. Consequently, any systematic error in the adopted $T_{\rm e}({\rm O}^{+})$ or $T_{\rm e}({\rm O}^{++})$ propagates into the derived ionic abundances and, ultimately, into the total oxygen abundance. Because collisionally excited line emissivities depend exponentially on temperature, even modest biases in the inferred temperature can produce notable abundance errors.

Importantly, these errors are not random. At fixed excitation, the sign and magnitude of the bias depend on the location within the $t_2$--$t_3$ plane. In the high-$O_{32}$ regime, $t_2$ tends to be higher than $t_3$ at low and intermediate $t_3$, but lower at the highest temperatures where the saturation branch develops. Consequently, the inferred O$^{+}$/H$^{+}$ abundance may be either underestimated or overestimated depending on the physical conditions. The resulting abundance bias therefore varies systematically with both excitation and electron temperature (and, therefore, metallicity) rather than representing a simple random uncertainty. Consequently, analyses based on classical one-dimensional $t_2$--$t_3$ relations may introduce spurious trends in O$^{+}$/H$^{+}$ and, in some cases, in the total oxygen abundance when applied to samples spanning a broad range of excitation conditions. Our results are complementary to those of \citet{Yates2020}, who investigated the distribution of H\,\textsc{ii} regions and galaxies in the $t_2$--$t_3$ plane and demonstrated that departures from the classical relation can introduce systematic biases in direct-method abundance determinations. Here we show that variations in the ionization state provide a physical interpretation for part of the observed distribution.

This issue is particularly relevant for highly excited star-forming systems, such as Green Pea galaxies, Lyman continuum leakers, and early galaxies observed with \textit{JWST}, which commonly exhibit higher ionization parameters, harder radiation fields, and more extreme O32 ratios than local \ion{H}{II} regions. Such systems occupy precisely the region of parameter space where the literature one-dimensional relations perform least well and where the new calibrations derived here are expected to provide a significant improvement, to be tested in Paper~II.

Our results suggest that direct-method chemical abundance scales calibrated primarily on nearby \ion{H}{II} regions with intermediate electron temperatures, where auroral lines are the strongest, cannot automatically be assumed to remain unbiased when applied to galaxies occupying more extreme areas of parameter space. Incorporating ionization-sensitive observables such as O32 or S2S3 provides a physically motivated way to account for these differences and reduce systematic uncertainties in abundance determinations. More broadly, our results support a transition from one-parameter temperature relations toward multi-parameter descriptions that explicitly account for the ionization state of the nebula and other parameters such as dust abundance. This is especially important for studying spatial variations of physical parameters in star-forming galaxies, where small biases may cause misleading conclusions. 

An important next step will be to extend observational constraints into areas of parameter space that remain sparsely sampled by current datasets, particularly at the highest and lowest electron temperatures and at extreme O32 values. High-precision auroral-line measurements in these regimes would provide a critical test of the ionization-dependent behavior predicted by the models and help establish the extent to which the 2D temperature relation derived here applies to real \ion{H}{II} regions.

\section{Conclusions}
\label{section:Summary}

We computed a large homogeneous grid of \textsc{Cloudy}~C23.01 photoionization
models of \ion{H}{ii} regions. The grid spans a broad range of metallicity, ionization parameter, N/O ratio, electron density, and age of the ionizing stellar population. We used it to investigate the relation between the electron temperatures in the high- and low-ionization zones. Ion-weighted temperatures, which characterize the thermal structure of the nebula, were compared with temperatures inferred from auroral-to-nebular line ratios, which are the quantities accessible to observations. Our main conclusions are as follows:

\begin{enumerate}

\item The classical assumption of a unique correspondence between $t_2$ and $t_3$ is not supported by our model grid. The relation is intrinsically non-linear and exhibits systematic secondary dependencies.

\item The dominant secondary parameter is the ionization state of the nebula. At fixed
$T_{\rm e}({\rm O}^{++})$, models separate systematically according to O32, S2S3 parameters, or the ionization parameter $U$. The hardness of the ionizing radiation field introduces a weaker additional dependence: models ionized by 4--6~Myr stellar populations influenced by WR stars exhibit larger temperature scatter than models ionized by younger stellar populations. Dust is the dominant remaining systematic uncertainty in the calibration at high temperatures. The grain abundance therefore tilts the relation rather than displacing it, by an amount that exceeds the residual scatter of the two-dimensional fit.

\item The observable O32 or S2S3 ratios provide an effective ionization-sensitive proxy. A two-dimensional calibration reduces the residual scatter from $\sim$3--4\% for the one-dimensional relation to $\sim$1--2\%, depending on stellar-population age. 

\item Line-derived and ion-weighted temperatures are not identical quantities. For most models they differ by less than a few percent. The difference reaches up to $\sim$20\% only in the coldest and most highly ionized systems, which are rarely encountered in observed \ion{H}{ii} regions. Among the diagnostics considered here, $T_{\rm e}([\ion{N}{ii}])$ provides a robust observational tracer of $T_{\rm e}({\rm O}^{+})$. $T_{\rm e}([\ion{S}{iii}])$ reproduces $T_{\rm e}({\rm O}^{++})$ over most of the grid, but degrades in the same cold, highly ionized corner.

\item One-dimensional $t_2$--$t_3$ relations can introduce systematic O/H biases in the high-ionization regime. The two-dimensional calibration is expected to reduce these biases. It also offers a natural explanation for the shallower $T_{\rm e}([\ion{O}{ii}])$--$T_{\rm e}([\ion{O}{iii}])$ slopes reported in recent high-redshift studies, which follow from the preferential sampling of high-temperature, high-O32 conditions and do not require any fundamental evolution of nebular physics.

\item The highest- and lowest-temperature, high-O32 regions of parameter space remain sparsely sampled observationally. Deep \textit{JWST}/NIRSpec spectroscopy of high-O32 systems and large auroral-line compilations of local metal-poor galaxies can now populate these regimes. Such measurements would provide a critical test of the ionization-dependent behavior predicted by the models.

\end{enumerate}

This paper is the first in a series aimed at establishing physically motivated electron-temperature relations for star-forming regions, from the local Universe to high redshift. In Paper~II we apply the relations derived here to local and high-redshift galaxy samples.

\begin{acknowledgements}
We are grateful to the referee for the constructive comments and suggestions that helped us improve the paper.
IAZ acknowledges funding from the Deutsche Forschungsgemeinschaft (DFG; German Research Foundation)---project-ID 550945879.
BPD acknowledges support from INAF RF2024 Large Grant "UNDUST: UNveiling the Dawn of the Universe with JWST"; INAF RF2024 GO Grant ”Revealing the nature of bright galaxies at cosmic dawn with deep JWST spectroscopy”
EPM acknowledges financial support from the Spanish Ministerio de Ciencia e Innovaci\'on and from the AEI  “Center of Excellence Severo Ochoa” award to the IAA (SEV-2017-0709).
The authors acknowledge support by the High Performance and Cloud Computing Group at the Zentrum für Datenverarbeitung of the University of Tübingen, the state of Baden-Württemberg through bwHPC and the German Research Foundation (DFG) for funding under "Project number 455787709" (bwForCluster BinAC 2).
\end{acknowledgements}

\bibliography{reference}

\clearpage
\onecolumn
\begin{appendix}

\section{Sensitivity to the atomic dataset}
\label{app:atomic}

\begin{figure}[H]
\resizebox{1.00\hsize}{!}{\includegraphics[angle=000]{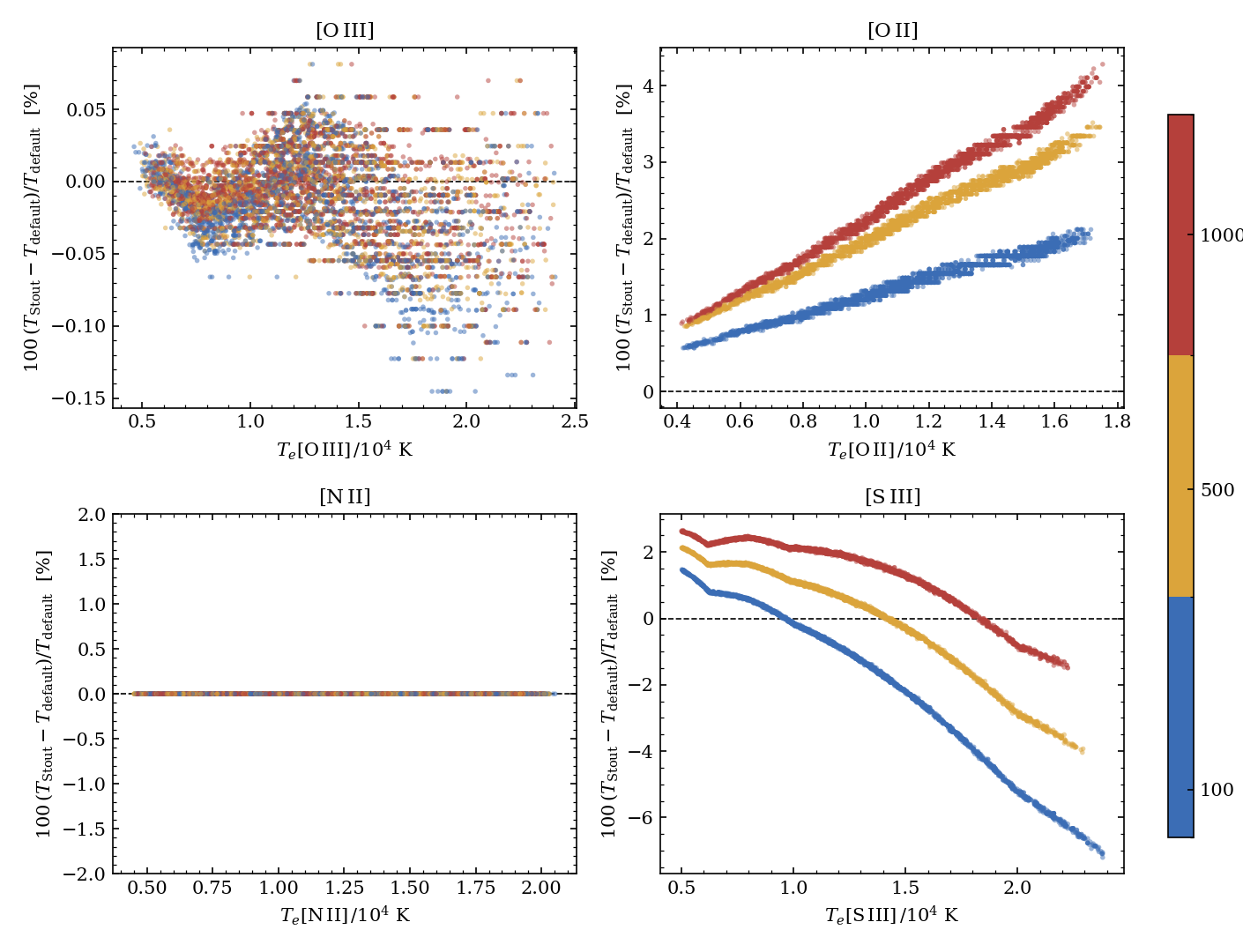}}
\caption{Relative difference between line-derived electron temperatures computed with the default \textsc{PyNeb} atomic data and with \textsc{Cloudy}'s Stout dataset, as a function of $T_{\rm e}$ (POPSTAR grid, 1--6~Myr), for [\ion{O}{iii}], [\ion{O}{ii}], [\ion{N}{ii}] and [\ion{S}{iii}]. Points are color-coded by electron density. The [\ion{N}{ii}] panel is identically zero because the two codes use the same underlying N$^{+}$ atomic data.}
\label{fig:atomic}
\end{figure}

All line-derived temperatures in this work are computed with \textsc{PyNeb} using, whenever available, the same atomic data adopted by \textsc{Cloudy}~C23.01 (the Stout database). This ensures that the line-derived and ion-weighted temperatures are based on a consistent set of atomic data. Since most observational studies instead use the default \textsc{PyNeb} atomic dataset (e.g., PYNEB\_23\_01), we assess the impact of this choice by recomputing all four temperature diagnostics over the full POPSTAR grid with both datasets.

Figure~\ref{fig:atomic} shows the resulting relative differences as a function of temperature, color-coded by electron density. For the high-ionization diagnostic defining the $t_3$, $T_{\rm e}([\ion{O}{iii}])$, the two datasets agree to better than 0.05\% over the entire density range. The calibrations presented in this paper may therefore be applied using either dataset without introducing any significant systematic error. 
However, $T_{\rm e}([\ion{O}{ii}])$ exhibits a measurable offset, increasing monotonically with density from $\sim1$\% to $\sim2-4$\%, depending on the electron density. Likewise, $T_{\rm e}([\ion{S}{iii}])$ differs by up to $\sim2-6$\%, depending on the electron density. $T_{\rm e}([\ion{N}{ii}])$ is identical by construction as \textsc{Cloudy} computes \ion{N}{ii} using its bundled \textsc{Chianti}~v10 data, whose collision strengths and transition probabilities are the same as those adopted by the default \textsc{PyNeb} dataset.

The discrepancies increase with temperature because of the different construction of the two datasets. Stout compiles atomic data from multiple literature sources to maximize the coverage of species and transitions \citep{Lykins2015}, whereas the default \textsc{PyNeb} dataset relies on a single reference for each atomic parameter and includes fewer transitions. Consequently, differences grow as higher energy levels become populated. The behavior of [\ion{S}{III}] is expected, since S$^{++}$ is among the least well-constrained ions. Similar discrepancies in \textsc{Cloudy} predictions for S$^{++}$ infrared lines have been reported by \citet{Perezdiaz2022}, particularly for harder ionizing spectra. By contrast, the [\ion{N}{II}] panel shows only a negligible offset because \textsc{Cloudy} and \textsc{PyNeb} adopt the same atomic data for this ion.

\section{Residuals organized by $\log(\mathrm{O32})$}
\label{app:logO32-residuals}

\begin{figure}[H]
\resizebox{0.85\hsize}{!}{\includegraphics[angle=000]{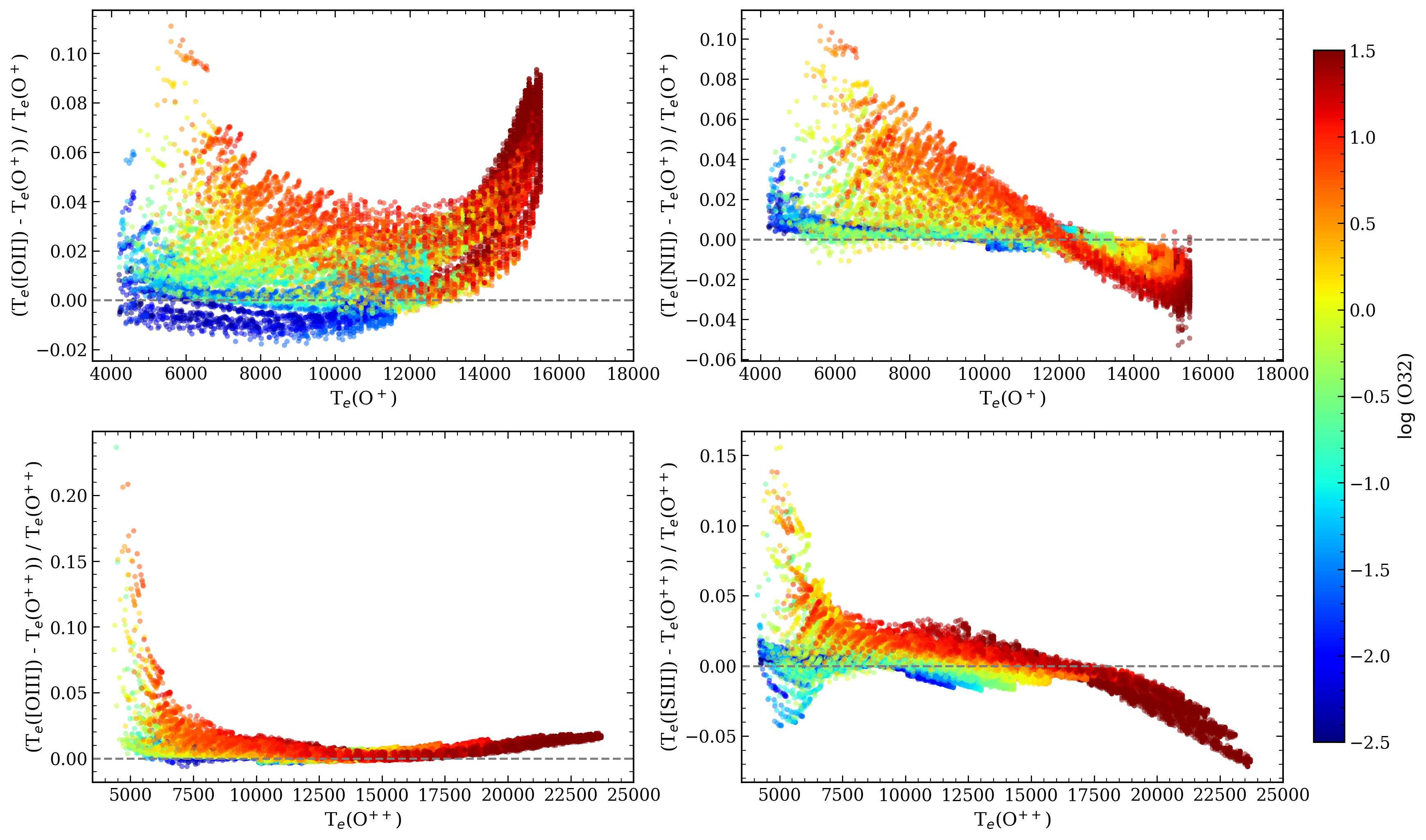}}
\caption{As Fig.~\ref{figure:Te-comparison-young} (1--3~Myr models) but color-coded by $\log(\mathrm{O32})$ instead of $\log U$.}
\label{figure:Te-comparison-young-logO32}
\end{figure}

\begin{figure}[H]
\resizebox{0.85\hsize}{!}{\includegraphics[angle=000]{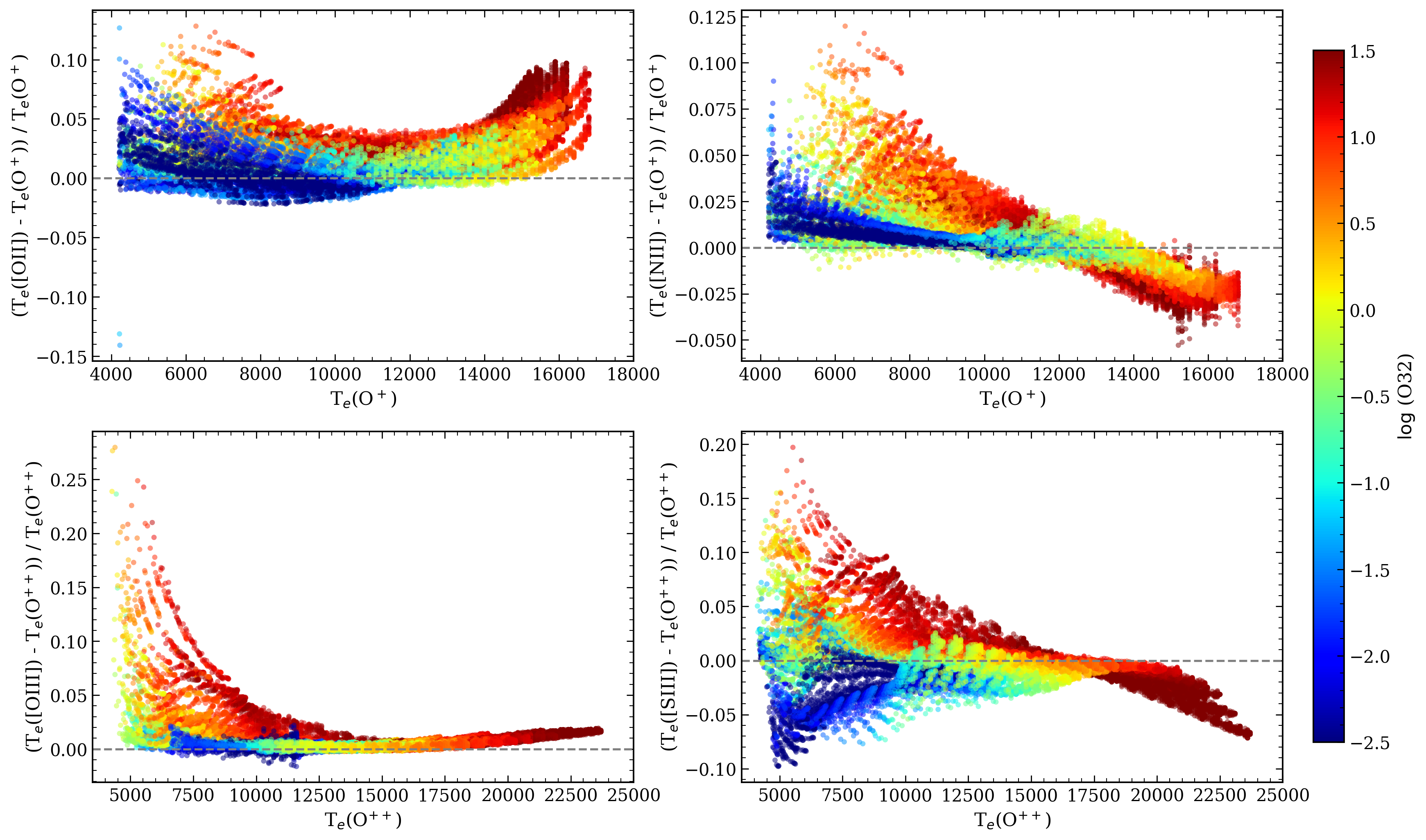}}
\caption{As Fig.~\ref{figure:Te-comparison-old} (full 1--6~Myr grid) but color-coded by $\log(\mathrm{O32})$ instead of $\log U$.}
\label{figure:Te-comparison-old-logO32}
\end{figure}

Figures~\ref{figure:Te-comparison-young} and \ref{figure:Te-comparison-old} show the relative differences between line-derived and ion-weighted temperatures color-coded by $\log U$. Because the ionization parameter is not directly observable, Figures~\ref{figure:Te-comparison-young-logO32} and \ref{figure:Te-comparison-old-logO32} reproduce the same diagnostic but color-coded by the observable $\log(\mathrm{O32})$. The structure of the residuals is essentially unchanged --- the same cold/high-excitation models that drive the high-$\log U$ outliers in Fig.~\ref{figure:Te-comparison-young} also drive the high-$\log(\mathrm{O32})$ outliers here --- showing that $\log(\mathrm{O32})$ is sensitive to the same ionization-state dependence that is used as the second variable of the two-dimensional fit in Sect.~\ref{sect:second-param}.

\end{appendix}

\end{document}